**A highly integrated, stand-alone photoelectrochemical device for large-scale solar hydrogen production**

Minoh Lee,[1*] Bugra Turan,[1] Jan-Philipp Becker,[1] Katharina Welter,[1] Benjamin Klingebiel,[1] Elmar Neumann,[2] Yoo Jung Sohn,[3] Tsvetelina Merdzhanova,[1] Thomas Kirchartz,[1,4] Friedhelm Finger,[1] Uwe Rau,[1] and Stefan Haas[1]

[1]IEK5 - Photovoltaik, Forschungszentrum Jülich GmbH, 52425 Jülich, Germany

[2]HNF, Forschungszentrum Jülich GmbH, 52425 Jülich, Germany

[3]IEK1, Forschungszentrum Jülich GmbH, 52425 Jülich, Germany

[4]Faculty of Engineering and CENIDE, University of Duisburg-Essen, Carl-Benz-Str. 199, 47057 Duisburg, Germany

E-mail: m.lee@fz-juelich.de

**Although photoelectrochemical water splitting is likely to be an important and powerful tool to provide environmentally friendly hydrogen, most developments in this field have been conducted on a laboratory scale so far. In order for the technology to make a sizeable impact on the energy transition, scaled up devices made of inexpensive and earth abundant materials must be developed. In this work, we demonstrate a scalable (64 $cm^2$ aperture area) artificial photoelectrochemical device composed of triple-junction thin-film silicon solar cells in conjunction with an electrodeposited bifunctional nickel iron molybdenum water splitting catalyst. Our device shows a solar to hydrogen efficiency of up to 4.67% (5.33% active area) without bias assistance and wire connection. Furthermore, gas separation was enabled by incorporating a membrane in a 3D printed device frame.**

## Introduction

Solar powered water electrolysis is an attractive way of storing chemical energy and of supplying green hydrogen for industry as hydrogen is generated by using unlimited and clean solar energy[1-4]. Among the various pathways for implementation of solar-hydrogen technologies, the usage of "photovoltaic-electrochemical (PV-EC) devices", which consist of a combination of the PV device as a power source and an electrocatalyst for splitting water, can be one of the economical options to generate sustainable hydrogen[5-9]. In such an integrated PV-EC device, charge carriers (e.g. electrons and holes) are generated in the PV part and transferred to the electrocatalyst to split water[10-16]. Although PV-EC hydrogen production has been regarded as a promising pathway in the field of hydrogen technology, most of the research has been carried out so far on a laboratory scale (≤ 1 $cm^2$). Problems that arise when scaling up PV-EC devices have hardly been tackled up to date. Our group has recently demonstrated a compact scalable PV-EC prototype device with an aperture area of 64 $cm^2$. The design uses two series connected thin-film silicon a-Si:H/μc-Si:H solar cells in conjunction with a nickel foam electrocatalyst[17]. Although the scalable concept represents a critical step towards industrial application, the ~3.9% solar-to-hydrogen (STH) efficiency of the modules must be further improved for successful commercial deployment[9]. In addition, the concept needs to be extended to include integrated gas management for the separate collection of hydrogen and oxygen because gas mixture might cause efficiency loss from the back reactions (e.g. hydrogen oxidation and oxygen reduction reaction) and danger of a gas explosion[18,19].

In the present study, we demonstrate further development of the PV-EC prototype presented in ref. 17, by tackling the challenges of both low STH efficiency and lack of gas management. To increase the efficiency of water splitting, we adapted the PV part of the device to the needs of the EC part by using efficient triple junction solar cells based on an a-Si:H/a-Si:H/μc-Si:H layer stack, as it was already been done for small (0.5 $cm^2$)[20] and large (64 $cm^2$)[21,22] scale devices. Additionally, we developed a bifunctional NiFeMo water splitting catalyst prepared by electrodeposition. The bifunctional catalysts enable the PV-EC system to get rid of the need for two different catalyst materials for the anode and cathode side, which offers the potential for lower costs in mass production[23-27]. The approach employing inexpensive earth-abundant materials together with a much simpler electrodeposition process would also lead to cost-effective production of the system, possibly resulting in one step closer to the commercialization[28-30]. Moreover, we have developed a 3D printed frame, on which the gas separation membrane, the PV module, and the catalysts can be mounted. The upscaled PV-EC device yields an STH efficiency of 4.67% with PV aperture area of 64 $cm^2$ (5.33% on the PV active area of 56 $cm^2$). In addition, the gas and electrolyte management of the modules including the gas separation was implemented in the frame of the modules that would prevent undesired side reaction and explosion. We expect that our advanced practical demonstration is an additional important step towards large-scale solar fuel production.

## Results

### Preparation of triple junction a-Si:H/a-Si:H/μc-Si:H solar cells

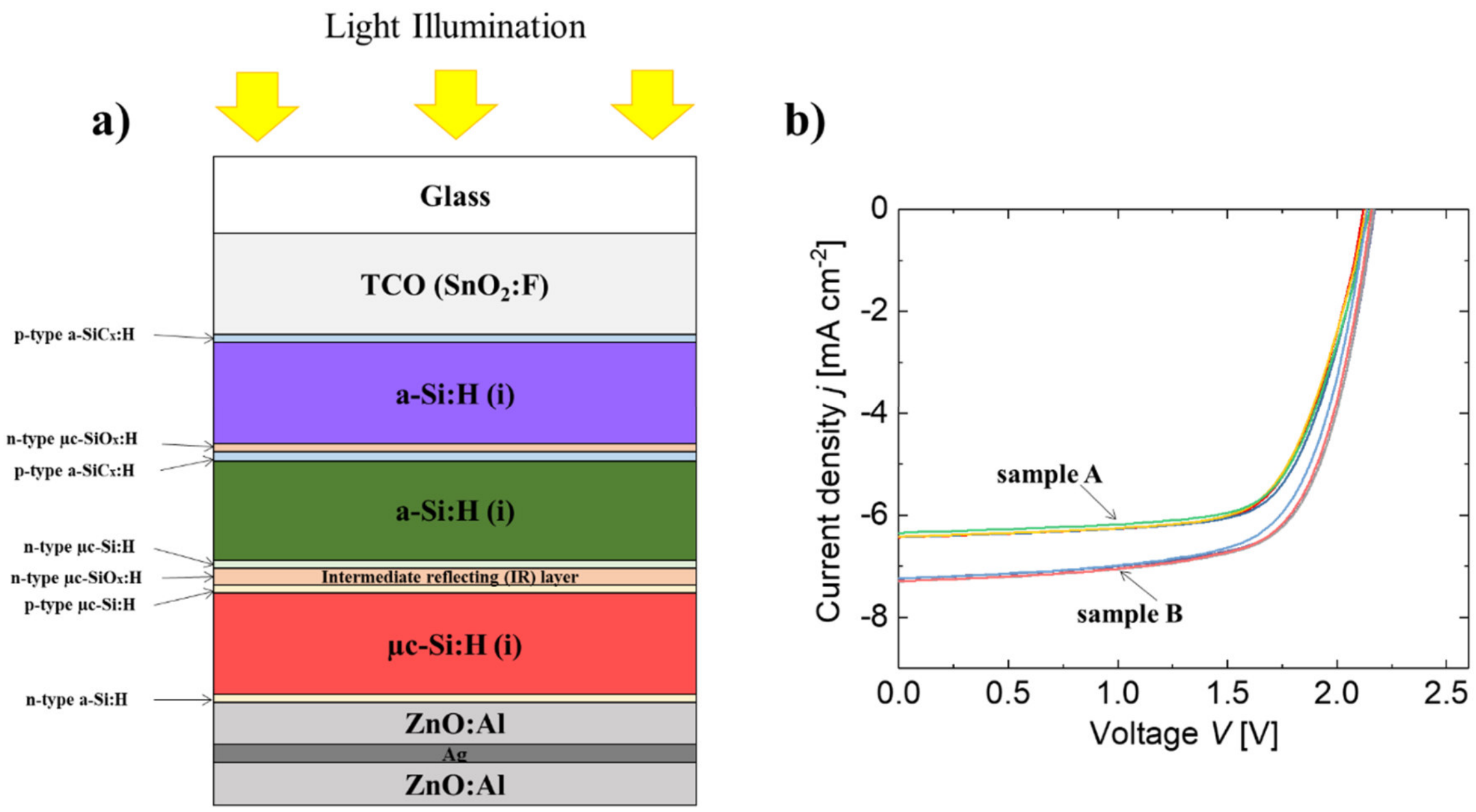

**Fig. 1 Characteristics of triple junction a-Si:H/a-Si:H/μc-Si:H solar cell. a,** Schematic image of the triple junction a-Si:H/a-Si:H/μc-Si:H solar cell in a superstrate configuration. **b,** *I-V* characteristics of samples prepared without intermediate reflecting (IR) layer (sample A) and with IR layer (sample B). After introducing an IR layer, the current density was increased from 6.41 (sample A; without IR layer) to 7.29 mA cm$^{-2}$ (sample B; with IR layer), resulting in an increase of the average value of solar to electricity conversion efficiency from 9.47 (sample A; without IR layer) to 10.83%. The samples have a size of 1.5 cm X 1.5 cm (2.25 cm$^2$ active area).

Many research groups have extensively investigated the application of multijunction cell structures consisting of vertically integrated amorphous (a-Si) and microcrystalline (μc-Si) layers for solar water splitting[17,20-22,31-47]. Such a vertically integrated multijunction structure provides an advantage of an open-circuit voltage in a wide range from 1.5 V to 2.8 V that is high enough to split water[20,22] (thermodynamic water splitting potential; $\Delta E_{\mathrm{themo}}$ = 1.23 V). Therefore, the multijunction structure doesn't need a lateral series connection of subcells which is a general approach to generate affordable voltage in the PV based spontaneous water splitting system[12-14,48]. In the present work, we have used triple junction a-Si:H/a-Si:H/μc-Si:H solar cell for the integrated solar water splitting device since a previous report shows best STH efficiency of 9.5% for thin-film silicon solar cells in this configuration due to maximum utilization of the sun spectrum under the constraint of the need for voltages > 1.6V[20].

Fig. 1a shows the schematic structure of a triple junction solar cell as investigated in the present study. We prepared two types of samples, 'sample A' without intermediate reflecting (IR) layer and 'sample B' with n-type μc-SiO$_x$:H IR layer between the middle and the bottom cell in order to reduce current mismatch between the subcells and thus increase the total current of the device[20,49,50]. The *I-V* characteristic of the solar cells is shown in Fig. 1b (PV parameters can be found in Supplementary Table 1). The average value of solar to electricity conversion efficiency of the 'sample B' (10.83%) could be increased compared to that of 'sample A' (9.47%). A significant increase in the current density (from 6.41 to 7.29 mA $cm^{-2}$) seems to be the main reason for the enhancement of cell efficiency by introducing an IR layer. The open-circuit voltage $V_{oc}$ of the devices is almost identical (2.14 V vs. 2.16 V). In conclusion, we expect to have higher STH efficiency in an integrated PV-EC device due to the higher current density of the triple-junction solar cell including IR layer[12,21].

### Development of electrodeposited bifunctional NiFeMo water splitting catalyst

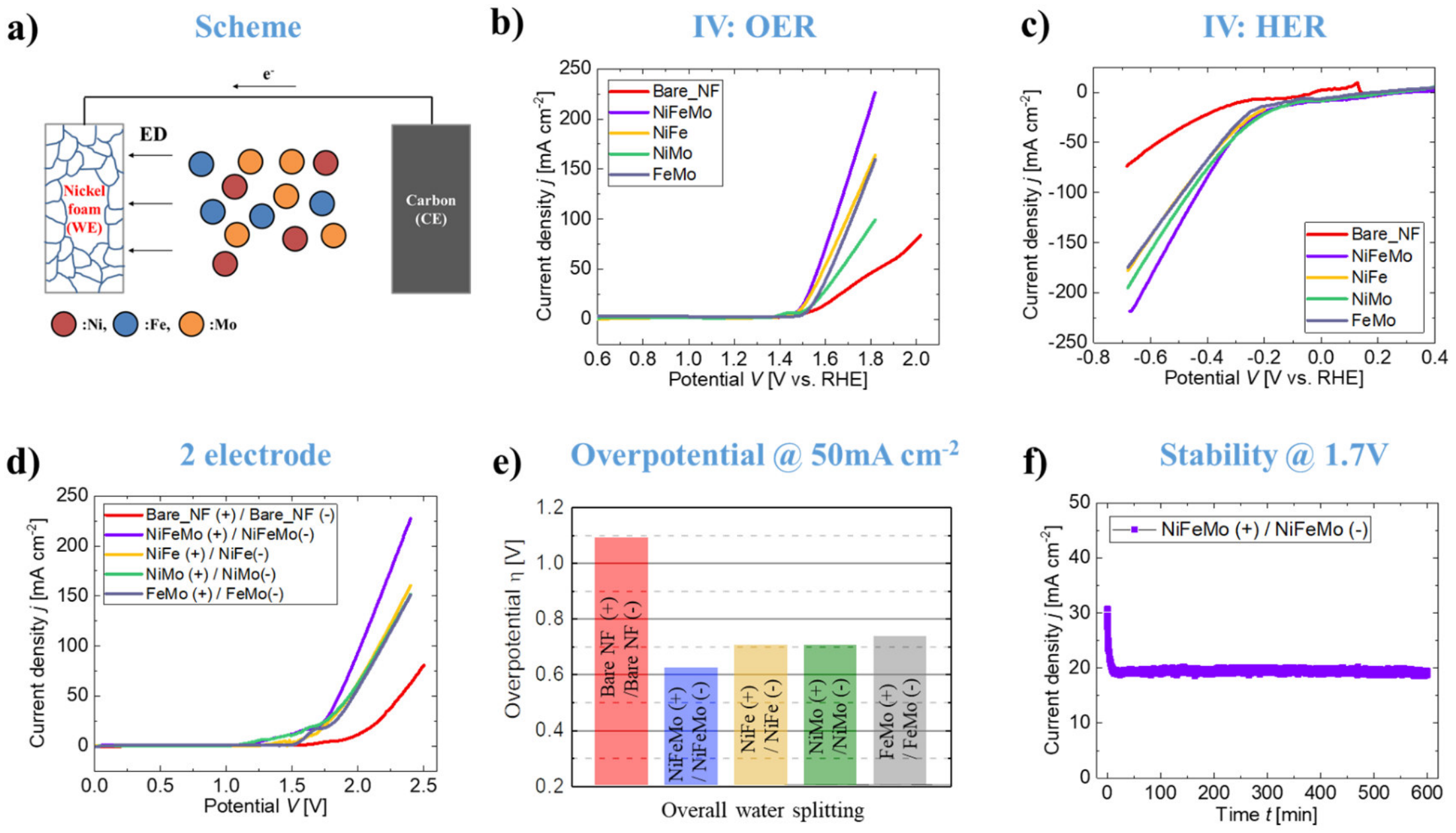

**Fig. 2 Electrochemical characterization of bifunctional NiFeMo electrocatalyst. a,** Schematic illustration of the electrodeposition method for the preparation of NiFeMo catalyst. ED indicates electrodeposition. **b-d,** Linear sweep voltammetry (LSV) curves of Bare_NF (bare nickel foam), NiFeMo, NiFe, NiMo, and FeMo for different reaction modes: **b,** Oxygen evolution reaction (OER), **c,** Hydrogen evolution reaction (HER), **d,** Overall water splitting using each material as a bifunctional electrocatalyst in a two-electrode configuration. Notably, all the catalysts are prepared on NF. **e,** Comparison of overpotential with prepared samples at a current density $J$ = 50 mA $cm^{-2}$. **f,** Chronoamperometry measurement of water splitting employing NiFeMo as a bifunctional catalyst in a two-electrode system under an applied voltage of 1.7 V. All tests were conducted in 1.0M KOH at room temperature.

The water electrolysis process consists of two half-reactions, the oxygen evolution reaction (OER) and the hydrogen evolution reaction (HER). A voltage considerably above the minimum energy for water splitting (1.23V), is required to split water in a real system due to the multi proton-coupled electron transfer reactions[51,52]. Thus efficient catalysts are required to accelerate these slow electron-transfer processes during water splitting. Although noble metals and their compounds (e.g. Pt-group metals for HER and $IrO_x$, $RuO_x$ for OER) are typically used as the benchmark electrocatalysts[53,54], their high price and scarcity could lead to considerable obstacles for the large technological implementation of water electrolysis systems for energy storage[55,56]. In order to find alternatives to the use of precious metal based catalysts, considerable research has been undertaken into the development of electrocatalysts based on earth-abundant materials (e.g. Ni, Co, Mo etc. and their compounds/mixture)[28,29]. In this study, we successfully developed an electrodeposited bifunctional electrocatalyst consisting of a NiFeMo compound for alkaline water electrolysis, which was inspired by previous research[21,26,53,57-59]. A schematic illustration of the electrodeposition process is presented in Fig. 2a. Trimetal NiFeMo catalysts (or other bimetallic samples; NiFe, NiMo, FeMo) were prepared on a nickel foam (NF) substrate. We decided to use NF in this stage due to its high surface area[60]. The deposition takes place in an ammonia solution consisting of 2.4M $NiSO_4 \cdot 6H_2O$, 0.6M $FeSO_4 \cdot 6H_2O$, 0.2M $Na_2MoO_4 \cdot 2H_2O$, and 0.3M $Na_3C_6H_5O_7 \cdot 2H_2O$ at a continuous cathodic current of -160 mA $cm^{-2}$ in a two-electrode configuration. For comparison, bimetallic samples (NiFe, NiMo, FeMo) were prepared in a similar manner. More details are provided in the experimental section of the Supplementary Information. To investigate the electrocatalytic properties of the as-prepared samples and to find optimized preparation conditions, linear sweep voltammetry (LSV) measurements were performed with the different samples in a three-electrode (for OER and HER) and a two-electrode configuration (overall water splitting) at a scan rate of 10 mV $s^{-1}$.

First, we varied the electrodeposition time for the NiFeMo trimetal catalyst to find the optimal conditions for overall water splitting (see Supplementary Fig.1). Supplementary Fig. 1 shows the average overpotentials at a current density of 50 mA $cm^{-2}$ as a function of the electrodeposition time for (a) OER, (b) HER and (c) overall reaction. Although the optimal deposition times for each half reaction are different (3 min for OER (a) and 5 min for HER (b)) we selected samples with 5 min deposition time for the integration into PV-EC devices, since the overall reaction (c) performs best at 5 min. For the comparison, bimetallic samples (NiFe, NiMo, FeMo) were also prepared with a similar process; at the same time (5 min) with different precursor solution (see Supplementary Fig. 2 and Table 2). The current density was found to depend on the catalyst deposition time. This phenomenon can be elucidated by a catalyst loading-activity relationship[61]. In particular, a better (or similar) performance was found with increasing deposition time up to 5 min (see Supplementary Fig. 1), while the films with thickness above the optimum (deposition times > 5 mins) don’t guarantee higher activity than thinner films (deposition times ≤ 5 mins). Probably the increased series resistance in too thick films hinders charge transport during water splitting. Figs. 2b-d show the representative polarization curves for the different catalysts in the OER (Fig. 2b), HER (Fig. 2c) and overall reaction (Fig. 2d). Additionally, the overall reaction overpotential at a current density of 50 mA $cm^{-2}$ is plotted in Fig. 2e. Among all samples, trimetal NiFeMo exhibits the lowest overpotential value of 0.62 V, which is considerably lower than that of the bare sample (~1.1 V), but also slightly lower than those of the bimetallic samples (~0.7-0.74 V). A measurement of the stability (Fig. 2f) in a continued chronoamperometry test at 1.7 V for 10 h shows almost no degradation.

To analyse whether the water splitting reaction induces any changes in the microstructure and composition of the catalysts, we performed scanning electron microscopy (SEM), energy-dispersive X-ray spectroscopy (EDX), grazing incidence X-ray diffraction (GIXRD), and X-ray photoelectron spectroscopy (XPS) on as prepared samples and on samples after the water splitting reaction. Fig. 3a shows a large area SEM image and corresponding EDX elemental mapping in which all the elements are quite uniformly distributed over hundreds of micrometers. Interestingly, the different elemental distributions of Ni, Fe, and Mo are clearly observed for the samples before and after reactions (Figs. 3b,c). In the case of OER, the dominant distribution of Fe and Mo is clearly distinctive in EDX element mapping images. On the other hand, elemental Ni was found to be dominant after HER in which Fe and Mo seem to have disappeared (or suppressed). The rearrangement of the catalyst elements after both reactions can be elucidated by a field-induced cation migration or a dissolution-precipitation mechanism[62-64]. These results imply real active sites involved in water splitting reaction would be different from the as-prepared state while keeping the activities stable. GIXRD was analyzed as seen in Supplementary Fig. 3 for as-prepared NiFeMo and samples after both reactions (OER and HER). All the samples measured with an incident angle of 2°, and even with a significantly lower angle of 0.1°, show no distinct peaks apart from the Ni substrate peaks. The coexistence of Ni, Fe, and Mo was confirmed by surface-sensitive XPS measurement (Supplementary Figs. 4-7). All the elements were mostly found to be in a high oxidation state for the Ni (2+), Fe (3+), and Mo (6+), even in the as-prepared sample. Combining the XPS data with GIXRD results, we assume an amorphous structure of mixture Ni, Fe, and Mo is distributed in the samples. The amorphous phase might be one of the reasons for enhancing the performance of the water splitting reaction since numerous defects could become active sites in the reaction[30,65-69]. Further investigations would be required to understand the detailed relationship between mixed metals and their oxidation states, which would be required to reach firm conclusions about the nature of the active sites involved in the water splitting reaction.

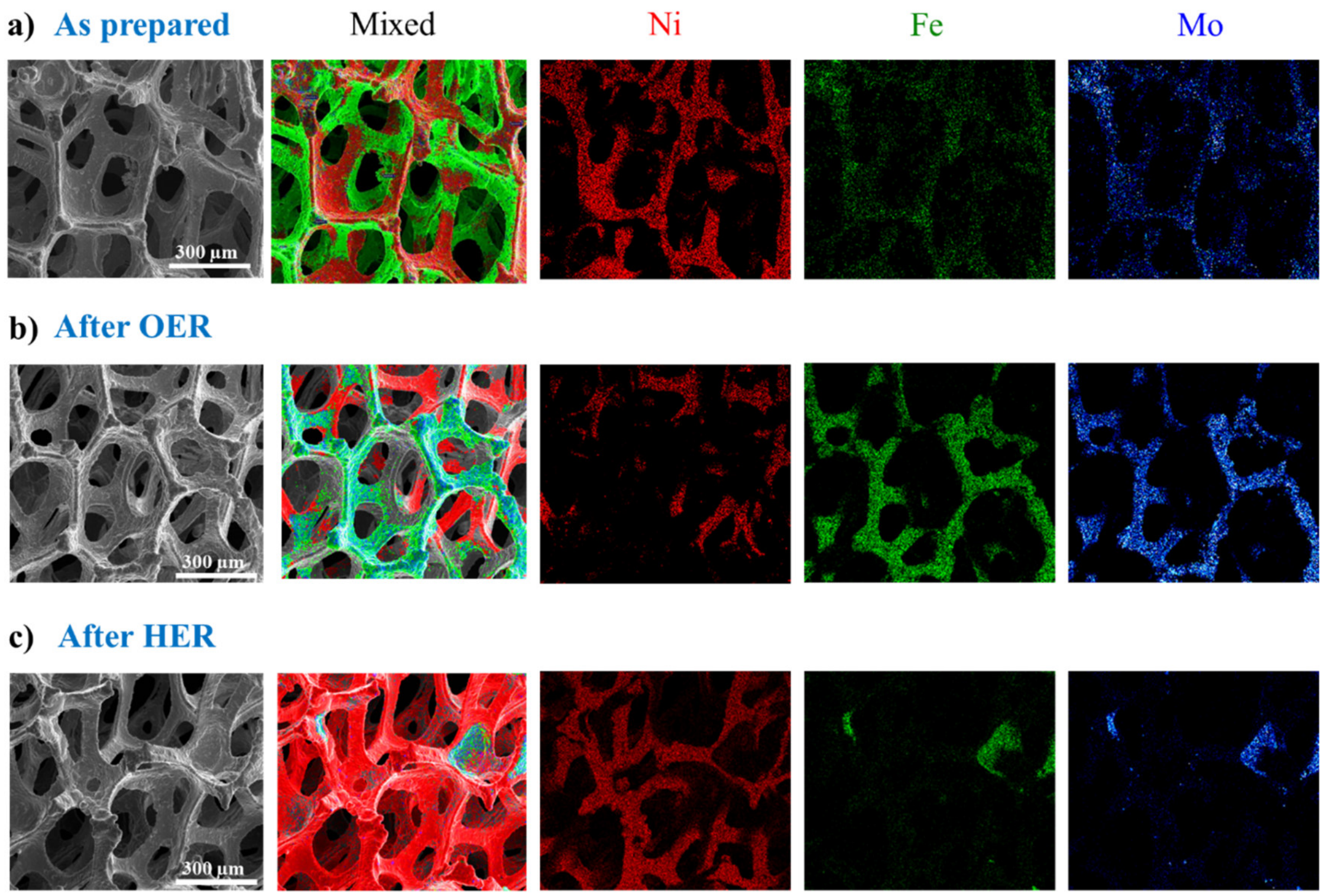

**Fig. 3 Comparison of the composition of bifunctional NiFeMo catalysts; a,** as prepared, **b,** after OER and **c,** after HER. The arrangement of images in each subsection is in the following order (from left to right); SEM image, EDS images for the mixture, Ni, Fe, and Mo. The rearrangement of the catalyst elements was observed after water splitting reaction, implying real active sites in the water splitting might be different compared to as-prepared states.

**Wired PV-EC device on small area (PV: 0.5 cm² as the photoactive area, EC: both 0.5 cm²)**

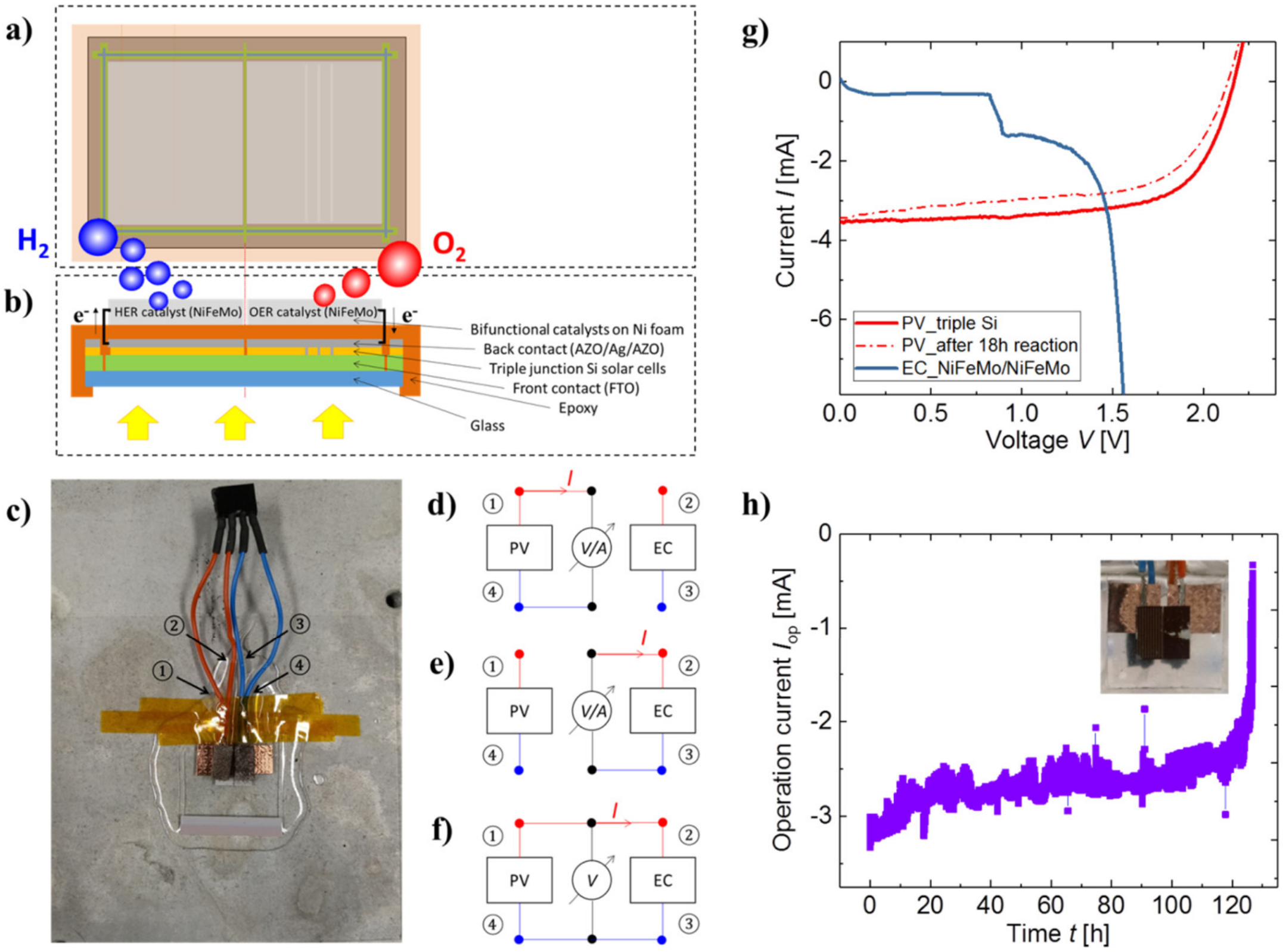

**Fig. 4 Wired PV-EC device on small area (PV: 0.5 cm² as the photoactive area, EC: both 0.5 cm²).** The schematic **a,** plain and **b,** cross-sectional view of PV-EC device and **c,** corresponding photograph of the lab scaled device where bifunctional NiFeMo catalysts (on both OER and HER side) were placed side-by-side on the back side of the triple junction Si solar cell. AZO and FTO indicate aluminum doped ZnO (ZnO:Al) and fluorine doped tin oxide, respectively. Block diagram of the PV-EC device as described in Fig. 4c and different operation modes for **d,** PV, **e,** EC, and **f,** combined PV-EC. Characterization of small area PV-EC device. **g,** *I-V* characteristics of PV and EC cell in a two-electrode system. **h,** Chronoamperometry of PV-EC device under bias-free (V=0) condition. Inset image represents a photograph of PV-EC device after continuous ~127 h reaction. All tests were conducted in 1.0M KOH at room temperature.

Before the development of the upscaled device, small area PV-EC devices were fabricated, which consist of one base unit of the upscaled device. A triple junction a-Si:H/a-Si:H/μc-Si:H solar cell and a bifunctional NiFeMo water splitting catalyst were combined in a lateral side-by-side configuration for the base unit. The schematics in different views (plain (Fig. 4a) and cross-sectional (Fig. 4b)) and photograph (Fig. 4c) of the integrated PV-EC device are provided in Figs. 4a-c, respectively. Superstrate configuration of solar cells provides a better light absorbing path without inevitable loss from interruption of catalysts and bubbles which are issues that commonly used water splitting systems are facing[45,70,71]. Wires (used here to measure the current and voltages of PV and EC, separately) were connected to the contacts of the PV cell by fixing them with copper tape and to the contacts of the EC cell by soldering with tin. We made separate contacts of the PV part and the EC part to allow for individual measurement of the I/V-characteristics. A transparent insulating epoxy (Loctite 9483) was used in order to protect the PV cell from the electrolyte and additionally decouple PV and EC elements. Figs. 4d-f show the illustration images of the block diagram with the different operation mode of the PV-EC device, thereby, both current and voltage can be measured in this configuration.

The performance of an integrated PV-EC device fixed into a well-sealed polyether ether ketone (PEEK) device holder was evaluated (see Supplementary Fig. 8b). The individual *I-V* curves of PV and EC are shown in Fig. 4g. The PV delivers an open-circuit voltage $V_{oc}$ of about 2.16 V and a voltage at the maximum power point $V_{mpp}$ of about 1.72 V, which is sufficiently high to enable bias-free water splitting. The intersection point, where both *I-V* curves of PV and EC meet, is considered as the operation current of the integrated PV-EC device without any assistance of bias. The intersection point is located at a current of -3.2 mA, corresponding to an active area STH efficiency of 7.87%. Operation current of an integrated PV-EC device was continuously monitored over 127 hours and plotted in Fig. 4h. The average STH efficiency for the first 1 h was determined to be 7.72%, which is in good agreement with the predicted value of 7.87%. The performance gradually degraded over 18 h. After removing the electrolyte, the *I-V* characteristic of the PV part was measured. Although both current and voltage of the PV part are not far from the original values, a significant reduction of the fill factor was observed (See the Fig. 4g), which can be linked to the light instability of amorphous thin-film Si (Staebler-Wronski effect)[72]. The measurement was resumed with a refreshed electrolyte, operation current was partially recovered and then gradually decreased until the device failed. Around 6.4% STH efficiency was achieved after 100 hours (calculated from average current for 1h; 99~100h) with 17% decrease from the initial value. Previously it was shown that ~100h continuous light illumination induces a degradation of ~12% in solar to electricity conversion efficiency of triple-junction a-Si:H/a-Si:H/μc-Si:H solar cell[73], which can be a main factor for the degradation of the integrated PV-EC device. Additionally, an increase in the coupling losses (e.g. wire and contact resistances) or partial degradation of the catalysts during the reaction may lead to the STH efficiency decrease. The device fully failed after around 127 h. Probably the epoxy was not stable for such a prolonged time, enabling hydroxide ion permeation to the silicon layer (see the inset in Fig. 4h).

## Wireless PV-EC device in a prototype scale (PV: 64 cm$^2$ aperture area, EC: both 26.1 cm$^2$)

The upscaled PV-EC device concept was realized in a 3D printed frame made of Digital ABS Plus material[74], which acts as a support for the PV module, the catalysts, and the membrane (see Fig. 5b). The use of the membrane inhibits potential losses from back reaction such as hydrogen oxidation and oxygen reduction reaction[18]. The risk of explosion from the accumulation of gas mixture can also be prevented by introducing a membrane[18]. In addition, gas management is implemented in the frame by several gas pipes. Both chamber types are additionally connected to the outside to allow a separate collection of the gases. Furthermore, pipes at the bottom of the frame can be used to replace used electrolyte. For the upscaled device, according to the previous approaches[21,22], we introduced nickel sheets (NS) as a substrate for the catalyst in contrast to the NF used for the catalyst development and the small area devices. While NS has a drawback of smaller surface area, it has the advantage of being impermeable. Thus the PV module is completely protected against corrosion from the alkaline electrolyte (see Fig. 5a). In addition, the present device can be made without wire connection between the PV and EC part which allows the device to be made simpler with flexible construction design[11,75]. Detailed information in terms of preparation procedure and dimensions can be found in the Supplemental Information.

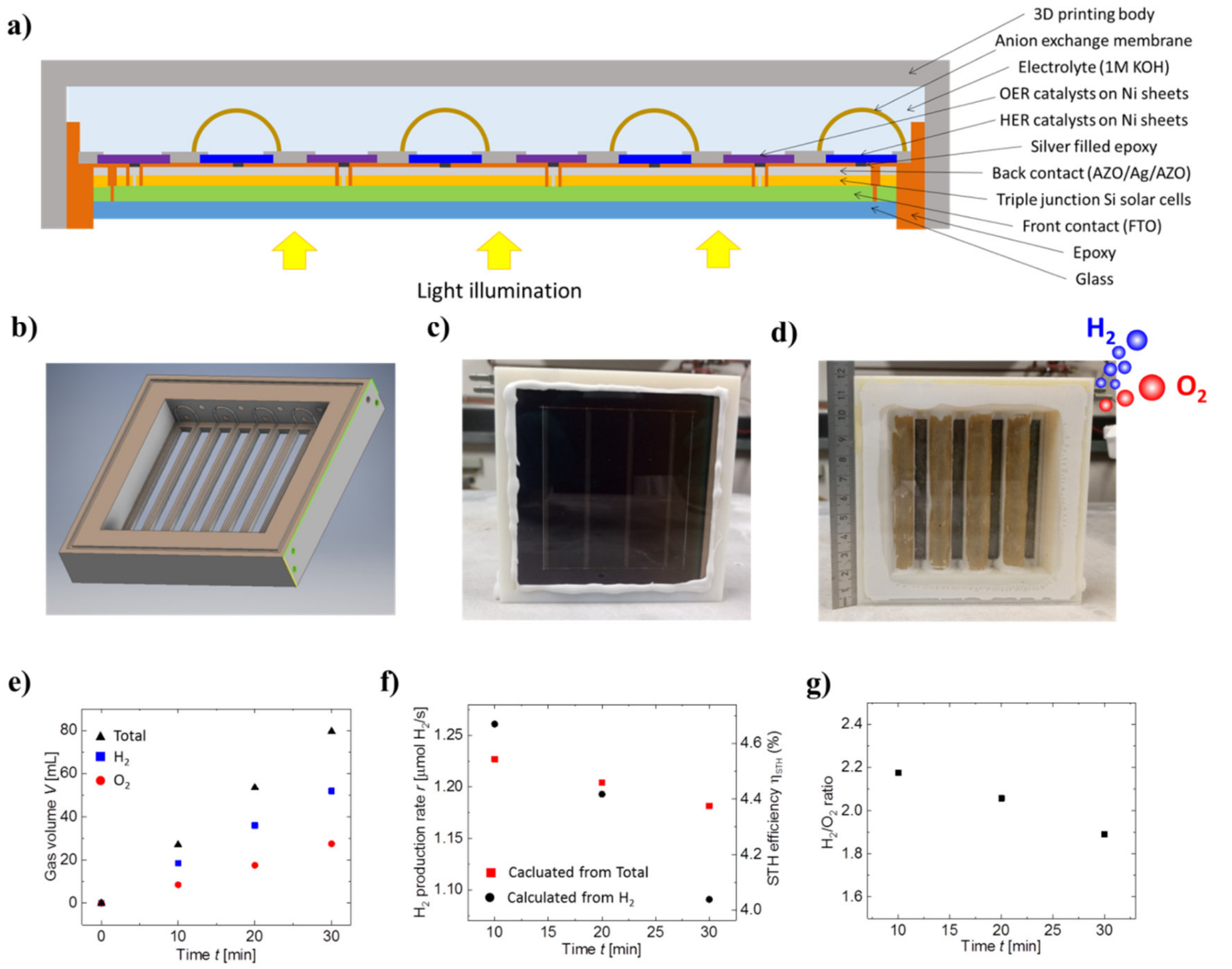

**Fig. 5 Wireless PV-EC device in a prototype scale (PV: 64 $cm^2$ aperture area, EC: both 26.1 $cm^2$). a,** Schematic illustration of an upscaled PV-EC devices as used in this study. AZO and FTO indicate aluminum doped ZnO (ZnO:Al) and fluorine doped tin oxide, respectively. **b,** Drawing of the 3D printed frame used to hold the catalysts, the solar module, and the membranes. Additionally, the frame is used for gas management. **c-d,** Photographs of upscaled PV-EC prototype device; **c,** front side view (PV side), **d,** the back side view (EC side) after integration of catalyst, membrane, and cover glass. **e-g,** Characterization of upscaled PV-EC device (direct gas measurement was conducted to evaluate STH efficiency instead of monitoring the operation current ($I_{op}$) since $I_{op}$ cannot be measured in the absence of wires); **e,** Collected gas ($H_2$, $O_2$ and total) volume, **f,** corresponding $H_2$ production rate and solar to hydrogen (STH) efficiency, and **g,** Ratio of $H_2$ to $O_2$ as a function of time.

Fig. 5a shows a schematic illustration of the upscaled PV-EC device. The base unit, which was introduced in the section of the small scale device, is continuously repeated to cover larger areas. Seven adjoining base units consisting of triple junction a-Si:H/a-Si:H/μc-Si:H solar cells and bifunctional NiFeMo water splitting catalyst were used. The upscaled wireless PV-EC device cannot be explored by measuring the current for the calculation of STH efficiency. To evaluate the performance of the device the produced amount of hydrogen and oxygen was monitored over time instead of monitoring the current. Fig. 5e shows the collected gas volume over time for $H_2$, $O_2$, and for the total volume. The upscaled PV-EC device yields a STH efficiency of up to 4.67% (5.33% active area) for the first 10 min (volume change was monitored after 5 min stabilization). Fig. 5f shows the $H_2$ production rate and STH efficiency calculated by both total and $H_2$ gas. It is noteworthy, that the STH efficiency calculated based on the total gas rate decreases much slower than that calculated based on the $H_2$ production rate. This is a clear indication for a gas crossover between the OER and the HER chambers (some $H_2$ moves to $O_2$ part). The crossover might be a result of an imperfect working anion exchange membrane due to the curved application and/or instability in the applied epoxy which glues the membranes to the 3D printed frame. The ratio of $H_2$ to $O_2$ is shown in Fig. 5g. The gas ratio at the first 10 min was found to be ~2.18, instead of the ideal value of 2. This discrepancy can most likely be explained by the lower Faradaic efficiency of $O_2$ compared to $H_2$ because more complex (4 $H^+/e^-$ transfer) process in OER would lead to being less successful for the gas evolution rather than a pathway of 2 $H^+/e^-$ transfer process in HER[51,52,76]. Additionally, it can be assumed that the surface state of the catalyst does not reach a steady state (e.g. redox reaction, dissolution, electromigration, etc.) although we measured the gas volume after 5 min of stabilization[62,77]. The gradual decrease observed in the ratio (from ~2.18 to ~1.89) can also be linked to the gas cross over mentioned above. Unfortunately, after 30 minutes of operation the device failed due to a detachment of the plexiglass cover. Nevertheless, we realized an increase of efficiency of around 16.5% compared to our previous result (~3.9%)[17], where two a-Si:H/μc-Si:H tandem solar cells serially connected together with bare NF as the catalyst were used. The enhancement of the efficiency can be ascribed to a better matching of the characteristics of the PV part and the EC part. In addition, an increase of the total active area (56 $cm^2$ vs. 52.8 $cm^2$) for a given total aperture area (64 $cm^2$) by reducing the number of base units (reducing the number of inactive anode contacts) can be one of the reasons for the improvement of STH efficiency. Although the laser-generated anode connection has only a width of ~100 μm, we need a much wider area for the anode contact (~2 mm) due to the manual deposition of the metal filled epoxy, which connects the anode contact to the OER catalyst. Thus, there is still room for improvement of the STH efficiency by further optimizing the device manufacturing processes by using e.g. more precise deposition techniques like inkjet printing or computer controlled dispensing.

## Conclusion

In this study, we successfully introduced an upscalable photoelectrochemical device which incorporates gas separation and gas handling, and which is based on earth-abundant materials. As power source, triple junction thin-film silicon solar cells were used. This type of solar cells offers almost perfect voltage matching to highly active catalysts, without the need for lateral series connection or even external bias. By introducing an intermediate reflecting layer into the solar device the current density was increased from 6.41 to 7.29 mA $cm^{-2}$. To achieve high STH efficiencies, we additionally developed stable bifunctional (OER and HER) NiFeMo water splitting catalysts made of inexpensive earth-abundant materials that we prepared by electrodeposition. Besides being a cost-effective route of electrodeposition for the sample preparation, the bifunctional property can additionally lower the cost of a PV-EC system, since it can be used with one type of catalyst in the same electrolyte.

By combining the triple junction thin-film silicon solar cell with the bifunctional NiFeMo catalyst an integrated PV-EC device with an active area of 0.5 $cm^2$ was realized. The device had an initial STH efficiency of ~7.7%. After 100h of operation, the STH efficiency was reduced by 17% (STH efficiency from 7.72 to 6.40%). Here the efficiency decrease might result from a combination of the light induced degradation (Staebler Wronsky effect) of the amorphous silicon layers in the solar cell, increasing coupling losses (e.g. wire and contact resistances), and partial degradation of the catalysts during the water-splitting reaction.

Finally, an upscaled device with an aperture area of 64 cm² was realized. Here a 3D printed frame was used acting as a support for the PV module, the catalysts, and the membrane. In addition, gas management and gas handling were realized in the frame. The upscaled device yields an STH efficiency of 4.67% (5.33% for the active area) during the first 10 min under bias-free condition. Although the upscaled device failed after 30 minutes of operation, the new device concept can be an attractive solution for gas management and wireless design. Furthermore, we expect that better sealing materials will lead to an enormous enhancement in stability. It is important to mention that our concept is not limited to thin-film silicon PV, it is compatible with almost any other superstrate thin-film PV technology like emerging organic-inorganic halide perovskites. In addition, the concept we proposed here, could be used as a tool to electrochemically convert $CO_2$ into fuels or commodity chemicals.

## Reference

1 Lewis, N. S. & Nocera, D. G. Powering the planet: Chemical challenges in solar energy utilization. *Proc. Natl. Acad. Sci. U.S.A.* **103**, 15729-15735, (2006).

2 Walter, M. G. *et al.* Solar Water Splitting Cells. *Chem. Rev.* **110**, 6446-6473, (2010).

3 Cook, T. R. *et al.* Solar Energy Supply and Storage for the Legacy and Nonlegacy Worlds. *Chem. Rev.* **110**, 6474-6502, (2010).

4 Chu, S. *et al.* Roadmap on solar water splitting: current status and future prospects. *Nano Futures* **1**, 022001, (2017).

5 Reuß, M. *et al.* Solar hydrogen production: a bottom-up analysis of different photovoltaic–electrolysis pathways. *Sustainable Energy Fuels* **3**, 801-813, (2019).

6 Jacobsson, T. J., Fjällström, V., Edoff, M. & Edvinsson, T. Sustainable solar hydrogen production: from photoelectrochemical cells to PV-electrolyzers and back again. *Energy Environ. Sci.* **7**, 2056-2070, (2014).

7 Jacobsson, T. J., Fjällström, V., Edoff, M. & Edvinsson, T. CIGS based devices for solar hydrogen production spanning from PEC-cells to PV-electrolyzers: A comparison of efficiency, stability and device topology. *Sol. Energy Mater. Sol. Cells* **134**, 185-193, (2015).

8 Ager, J. W., Shaner, M. R., Walczak, K. A., Sharp, I. D. & Ardo, S. Experimental demonstrations of spontaneous, solar-driven photoelectrochemical water splitting. *Energy Environ. Sci.* **8**, 2811-2824, (2015).

9 Ardo, S. *et al.* Pathways to electrochemical solar-hydrogen technologies. *Energy Environ. Sci.* **11**, 2768-2783, (2018).

10 Khaselev, O. & Turner, J. A. A Monolithic Photovoltaic-Photoelectrochemical Device for Hydrogen Production via Water Splitting. *Science* **280**, 425-427, (1998).

11 Reece, S. Y. *et al.* Wireless Solar Water Splitting Using Silicon-Based Semiconductors and Earth-Abundant Catalysts. *Science* **334**, 645-648, (2011).

12 Cox, C. R., Lee, J. Z., Nocera, D. G. & Buonassisi, T. Ten-percent solar-to-fuel conversion with nonprecious materials. *Proc. Natl. Acad. Sci. U.S.A.* **111**, 14057-14061, (2014).

13 Jacobsson, T. J., Fjällström, V., Sahlberg, M., Edoff, M. & Edvinsson, T. A monolithic device for solar water splitting based on series interconnected thin film absorbers reaching over 10% solar-to-hydrogen efficiency. *Energy Environ. Sci.* **6**, 3676-3683, (2013).

14 Luo, J. *et al.* Water photolysis at 12.3% efficiency via perovskite photovoltaics and Earth-abundant catalysts. *Science* **345**, 1593-1596, (2014).

15 Jia, J. *et al.* Solar water splitting by photovoltaic-electrolysis with a solar-to-hydrogen efficiency over 30%. *Nat. Commun.* **7**, 13237, (2016).

16 Landman, A. *et al.* Photoelectrochemical water splitting in separate oxygen and hydrogen cells. *Nat. Mater.* **16**, 646, (2017).

17 Turan, B. *et al.* Upscaling of integrated photoelectrochemical water-splitting devices to large areas. *Nat. Commun.* **7**, 12681, (2016).

18 Modestino, M. A. *et al.* Robust production of purified H2 in a stable, self-regulating, and continuously operating solar fuel generator. *Energy Environ. Sci.* **7**, 297-301, (2014).

19 Berger, A., Segalman, R. A. & Newman, J. Material requirements for membrane separators in a water-splitting photoelectrochemical cell. *Energy Environ. Sci.* **7**, 1468-1476, (2014).
20 Urbain, F. *et al.* Multijunction Si photocathodes with tunable photovoltages from 2.0 V to 2.8 V for light induced water splitting. *Energy Environ. Sci.* **9**, 145-154, (2016).
21 Becker, J. P. *et al.* A modular device for large area integrated photoelectrochemical water-splitting as a versatile tool to evaluate photoabsorbers and catalysts. *J. Mater. Chem. A* **5**, 4818-4826, (2017).
22 Welter, K. *et al.* Catalysts from earth abundant materials in a scalable, stand-alone photovoltaic-electrochemical module for solar water splitting. *J. Mater. Chem. A* **6**, 15968-15976, (2018).
23 Han, N. *et al.* Nitrogen-doped tungsten carbide nanoarray as an efficient bifunctional electrocatalyst for water splitting in acid. *Nat. Commun.* **9**, 924, (2018).
24 Yu, F. *et al.* High-performance bifunctional porous non-noble metal phosphide catalyst for overall water splitting. *Nat. Commun.* **9**, 2551, (2018).
25 Wang, A.-L., Xu, H. & Li, G.-R. NiCoFe Layered Triple Hydroxides with Porous Structures as High-Performance Electrocatalysts for Overall Water Splitting. *ACS Energy Lett.* **1**, 445-453, (2016).
26 Qin, F. *et al.* Trimetallic NiFeMo for Overall Electrochemical Water Splitting with a Low Cell Voltage. *ACS Energy Lett.* **3**, 546-554, (2018).
27 Jamesh, M. I. Recent progress on earth abundant hydrogen evolution reaction and oxygen evolution reaction bifunctional electrocatalyst for overall water splitting in alkaline media. *J. Power Sources* **333**, 213-236, (2016).
28 McKone, J. R., Marinescu, S. C., Brunschwig, B. S., Winkler, J. R. & Gray, H. B. Earth-abundant hydrogen evolution electrocatalysts. *Chem. Sci.* **5**, 865-878, (2014).
29 Hunter, B. M., Gray, H. B. & Müller, A. M. Earth-Abundant Heterogeneous Water Oxidation Catalysts. *Chem. Rev.* **116**, 14120-14136, (2016).
30 Lu, X. & Zhao, C. Electrodeposition of hierarchically structured three-dimensional nickel–iron electrodes for efficient oxygen evolution at high current densities. *Nat. Commun.* **6**, 6616, (2015).
31 Delahoy, A. E., Gau, S. C., Murphy, O. J., Kapur, M. & Bockris, J. O. M. A one-unit photovoltaic electrolysis system based on a triple stack of amorphous silicon (pin) cells. *Int. J. Hydrog. Energy* **10**, 113-116, (1985).
32 Appleby, A. J. *et al.* An amorphous silicon-based one-unit photovoltaic electrolyzer. *Energy* **10**, 871-876, (1985).
33 Sakai, Y., Sugahara, S., Matsumura, M., Nakato, Y. & Tsubomura, H. Photoelectrochemical water splitting by tandem type and heterojunction amorphous silicon electrodes. *Can. J. Chem.* **66**, 1853-1856, (1988).
34 Lin, G. H., Kapur, M., Kainthla, R. C. & Bockris, J. O. M. One step method to produce hydrogen by a triple stack amorphous silicon solar cell. *Appl. Phys. Lett.* **55**, 386-387, (1989).
35 Gramaccioni, C., Selvaggi, A. & Galluzzi, F. Thin film multi-junction solar cell for water photoelectrolysis. *Electrochim. Acta* **38**, 111-113, (1993).
36 Rocheleau, R. E., Miller, E. L. & Misra, A. High-Efficiency Photoelectrochemical Hydrogen Production Using Multijunction Amorphous Silicon Photoelectrodes. *Energy Fuels* **12**, 3-10, (1998).
37 Khaselev, O., Bansal, A. & Turner, J. A. High-efficiency integrated multijunction photovoltaic/electrolysis systems for hydrogen production. *Int. J. Hydrog. Energy* **26**, 127-132, (2001).
38 Yamada, Y. *et al.* One chip photovoltaic water electrolysis device. *Int. J. Hydrog. Energy* **28**, 1167-1169, (2003).
39 Miller, E. L., Paluselli, D., Marsen, B. & Rocheleau, R. E. Development of reactively sputtered metal oxide films for hydrogen-producing hybrid multijunction photoelectrodes. *Sol. Energy Mater. Sol. Cells* **88**, 131-144, (2005).
40 Cristino, V. *et al.* Efficient solar water oxidation using photovoltaic devices functionalized with earth-abundant oxygen evolving catalysts. *Phys. Chem. Chem. Phys.* **15**, 13083-13092, (2013).
41 Han, L. *et al.* Efficient Water-Splitting Device Based on a Bismuth Vanadate Photoanode and Thin-Film Silicon Solar Cells. *ChemSusChem* **7**, 2832-2838, (2014).

42 Urbain, F. *et al.* a-Si:H/μc-Si:H tandem junction based photocathodes with high open-circuit voltage for efficient hydrogen production. *J. Mater. Res.* **29**, 2605-2614, (2014).
43 Ziegler, J. *et al.* Photoelectrochemical and Photovoltaic Characteristics of Amorphous-Silicon-Based Tandem Cells as Photocathodes for Water Splitting. *ChemPhysChem* **15**, 4026-4031, (2014).
44 Urbain, F. *et al.* Application and modeling of an integrated amorphous silicon tandem based device for solar water splitting. *Sol. Energy Mater. Sol. Cells* **140**, 275-280, (2015).
45 Bogdanoff, P. *et al.* Artificial Leaf for Water Splitting Based on a Triple-Junction Thin-Film Silicon Solar Cell and a PEDOT:PSS/Catalyst Blend. *Energy Technol.* **4**, 230-241, (2016).
46 Azarpira, A. *et al.* Optimized immobilization of ZnO:Co electrocatalysts realizes 5% efficiency in photo-assisted splitting of water. *J. Mater. Chem. A* **4**, 3082-3090, (2016).
47 Urbain, F. *et al.* Multilayered Hematite Nanowires with Thin-Film Silicon Photovoltaics in an All-Earth-Abundant Hybrid Tandem Device for Solar Water Splitting. *ChemSusChem* **12**, 1428-1436, (2019).
48 Schüttauf, J.-W. *et al.* Solar-to-Hydrogen Production at 14.2% Efficiency with Silicon Photovoltaics and Earth-Abundant Electrocatalysts. *J. Electrochem. Soc.* **163**, F1177-F1181, (2016).
49 Lambertz, A. *et al.* Microcrystalline silicon–oxygen alloys for application in silicon solar cells and modules. *Sol. Energy Mater. Sol. Cells* **119**, 134-143, (2013).
50 Kirner, S. *et al.* An improved silicon-oxide-based intermediate-reflector for micromorph solar cells. *Phys. Status Solidi C* **9**, 2145-2148, (2012).
51 Dau, H. *et al.* The Mechanism of Water Oxidation: From Electrolysis via Homogeneous to Biological Catalysis. *ChemCatChem* **2**, 724-761, (2010).
52 Roger, I., Shipman, M. A. & Symes, M. D. Earth-abundant catalysts for electrochemical and photoelectrochemical water splitting. *Nat. Rev. Chem.* **1**, 0003, (2017).
53 McCrory, C. C. L. *et al.* Benchmarking Hydrogen Evolving Reaction and Oxygen Evolving Reaction Electrocatalysts for Solar Water Splitting Devices. *J. Am. Chem. Soc.* **137**, 4347-4357, (2015).
54 McCrory, C. C. L., Jung, S., Peters, J. C. & Jaramillo, T. F. Benchmarking Heterogeneous Electrocatalysts for the Oxygen Evolution Reaction. *J. Am. Chem. Soc.* **135**, 16977-16987, (2013).
55 Song, F. *et al.* Transition Metal Oxides as Electrocatalysts for the Oxygen Evolution Reaction in Alkaline Solutions: An Application-Inspired Renaissance. *J. Am. Chem. Soc.* **140**, 7748-7759, (2018).
56 Mahmood, N. *et al.* Electrocatalysts for Hydrogen Evolution in Alkaline Electrolytes: Mechanisms, Challenges, and Prospective Solutions. *Adv. Sci.* **5**, 1700464, (2018).
57 Fan, C., Piron, D. L., Sleb, A. & Paradis, P. Study of Electrodeposited Nickel-Molybdenum, Nickel-Tungsten, Cobalt-Molybdenum, and Cobalt-Tungsten as Hydrogen Electrodes in Alkaline Water Electrolysis. *J. Electrochem. Soc.* **141**, 382-387, (1994).
58 Raj, I. A. & Vasu, K. I. Transition metal-based cathodes for hydrogen evolution in alkaline solution: Electrocatalysis on nickel-based ternary electrolytic codeposits. *J. Appl. Electrochem.* **22**, 471-477, (1992).
59 Jayalakshmi, M.*,* Kim, W. Y., Jung, K. D., & Joo, O. S. Electrochemical Characterization of Ni-Mo-Fe Composite Film in Alkali Solution. *Int. J. Electrochem. Sci.* **3**, 908-917 (2008).
60 Xiao, C., Li, Y., Lu, X. & Zhao, C. Bifunctional Porous NiFe/NiCo2O4/Ni Foam Electrodes with Triple Hierarchy and Double Synergies for Efficient Whole Cell Water Splitting. *Adv. Funct. Mater.* **26**, 3515-3523, (2016).
61 Morales-Guio, C. G., Liardet, L. & Hu, X. Oxidatively Electrodeposited Thin-Film Transition Metal (Oxy)hydroxides as Oxygen Evolution Catalysts. *J. Am. Chem. Soc.* **138**, 8946-8957, (2016).
62 Singh, J. P., Lu, T. M. & Wang, G. C. Field-induced cation migration in Cu oxide films by in situ scanning tunneling microscopy. *Appl. Phys. Lett.* **82**, 4672-4674, (2003).
63 Lee, C. B. *et al.* Electromigration effect of Ni electrodes on the resistive switching characteristics of NiO thin films. *Appl. Phys. Lett.* **91**, 082104, (2007).
64 Schäfer, H. *et al.* Stainless steel made to rust: a robust water-splitting catalyst with benchmark characteristics. *Energy Environ. Sci.* **8**, 2685-2697, (2015).
65 Smith, R. D. L. *et al.* Photochemical Route for Accessing Amorphous Metal Oxide Materials for Water Oxidation Catalysis. *Science* **340**, 60-63, (2013).
66 Yang, Y., Fei, H., Ruan, G., Xiang, C. & Tour, J. M. Efficient Electrocatalytic Oxygen Evolution on Amorphous Nickel–Cobalt Binary Oxide Nanoporous Layers. *ACS Nano* **8**, 9518-9523, (2014).
67 Bergmann, A. *et al.* Reversible amorphization and the catalytically active state of crystalline Co3O4 during oxygen evolution. *Nat. Commun.* **6**, 8625, (2015).
68 Morales-Guio, C. G. & Hu, X. Amorphous Molybdenum Sulfides as Hydrogen Evolution Catalysts. *Acc. Chem. Res.* **47**, 2671-2681, (2014).
69 Merki, D., Fierro, S., Vrubel, H. & Hu, X. Amorphous molybdenum sulfide films as catalysts for electrochemical hydrogen production in water. *Chem. Sci.* **2**, 1262-1267, (2011).
70 Morales-Guio, C. G. *et al.* An Optically Transparent Iron Nickel Oxide Catalyst for Solar Water Splitting. *J. Am. Chem. Soc.* **137**, 9927-9936, (2015).
71 Oh, S., Song, H. & Oh, J. An Optically and Electrochemically Decoupled Monolithic Photoelectrochemical Cell for High-Performance Solar-Driven Water Splitting. *Nano Lett.* **17**, 5416-5422, (2017).
72 Staebler, D. L. & Wronski, C. R. Reversible conductivity changes in discharge-produced amorphous Si. *Appl. Phys. Lett.* **31**, 292-294, (1977).
73 Urbain, F., Smirnov, V., Becker, J.-P. & Finger, F. Impact of Light-Induced Degradation on the Performance of Multijunction Thin-Film Silicon-Based Photoelectrochemical Water-Splitting Devices. *ACS Omega* **1**, 832-836, (2016).
74 https://www.stratasys.com/materials/search/digital-abs-plus
75 Kim, J. H. *et al.* Wireless Solar Water Splitting Device with Robust Cobalt-Catalyzed, Dual-Doped BiVO4 Photoanode and Perovskite Solar Cell in Tandem: A Dual Absorber Artificial Leaf. *ACS Nano* **9**, 11820-11829, (2015).
76 Zeng, K. & Zhang, D. Recent progress in alkaline water electrolysis for hydrogen production and applications. *Prog. Energy Combust. Sci.* **36**, 307-326, (2010).
77 Spöri, C., Kwan, J. T. H., Bonakdarpour, A., Wilkinson, D. P. & Strasser, P. The Stability Challenges of Oxygen Evolving Catalysts: Towards a Common Fundamental Understanding and Mitigation of Catalyst Degradation. *Angew. Chem. Int. Ed.* **56**, 5994-6021, (2017).

## Acknowledgements

We gratefully acknowledge financial support from the “PECSYS” project, which has received funding from the Fuel Cells and Hydrogen 2 Joint Undertaking under grant agreement No 735218. This joint Undertaking receives support by the European Union’s Horizon 2020 research and innovation programme and Hydrogen Europe and N. ERGHY. We thank the Initiative and Networking Fund of the Helmholtz Association for funding of the JOSEPH cluster system via the Helmholtz Energy Materials Characterization Platform (HEMCP). We thank J. Kirchhoff, G. Schöpe, C. Zahren, S. Kasper, H. Siekmann, A. Bauer, and A. Gerber for their contributions to this work.

## Author contributions

U.R., F.F. and S.H. developed the conceptual idea and supervised the work. M.L. performed the experiments, collected and analyzed the data, and prepared a draft of the manuscript. B.T., J.-P.B., K.W., B.K., E.N., Y.J.S., T.M. and T.K. performed the experimental measurements and analysis. All of the authors discussed the results and commented on the manuscript.

## Additional information

The authors declare no competing interests.

# Supplementary Information

## A highly integrated, stand-alone photoelectrochemical device for large-scale solar hydrogen production

Minoh Lee,[1*] Bugra Turan,[1] Jan-Philipp Becker,[1] Katharina Welter,[1] Benjamin Klingebiel,[1] Elmar Neumann,[2] Yoo Jung Sohn,[3] Tsvetelina Merdzhanova,[1] Thomas Kirchartz,[1] Friedhelm Finger,[1] Uwe Rau,[1] and Stefan Haas[1]

[1]IEK5 - Photovoltaik, Forschungszentrum Jülich GmbH, 52425 Jülich, Germany
[2]HNF, Forschungszentrum Jülich GmbH, 52425 Jülich, Germany
[3]IEK1, Forschungszentrum Jülich GmbH, 52425 Jülich, Germany
[4]Faculty of Engineering and CENIDE, University of Duisburg-Essen, Carl-Benz-Str. 199, 47057 Duisburg, Germany

E-mail: m.lee@fz-juelich.de

**Table of Contents**

## Experimental section

## Fabrication of solar cells

Triple junction Si solar cells (a-Si:H/a-Si:H/μc-Si:H) were prepared in a stacked p-i-n superstrate configuration on the F-doped $SnO_2$ ($SnO_2$:F) coated glass. All intrinsic, n- and p-type thin film Si layers were deposited by plasma enhanced chemical vapor deposition (PECVD) method. As back contact, a layered stack of Al-doped ZnO and Ag (ZnO:Al/Ag/ZnO:Al) were prepared via radio frequency magnetron sputtering. We applied laser processing at varying conditions in order to make a lateral arrangement of front and back contact of the solar cells. Additional information in terms of layer deposition and laser patterning can be found in some of our previous works[1-3].

## Preparation of electrocatalyst

All reagents were used without further purification. Ni foam (NF, in a size of 3 cm × 1 cm for comparison of their properties and 1 cm × 0.5 cm for the implementation into a small PV-EC device) and Ni sheet (NS, in a size of 8.15 cm × 0.8 cm) was used as the substrate to deposit catalytic active substances. All the water splitting catalysts in this study were prepared by electrodeposition method employing a Gamry Reference 600 potentiostat. Carbon rod was used as the counter electrode in a two-electrode setup. Prior to electrodeposition, both NF and NS substrates were cleaned with 3M HCl (3M) solution in a sonication bath to remove natural oxidation layer on the surface. Followed by, the substrates were cleaned with ethanol and rinsed with deionized (DI) water.

### Electrodeposition of NiFe catalysts

Electrodeposition was carried out in a 0.1M solution (in DI water) consisting of $Ni(NO_3)_2{\cdot}6H_2O$ (98% Alfa Aesar) and $FeCl_2{\cdot}4H_2O$ (Emsure) with a molar ratio of 8:2 (Ni:Fe). Before electrodeposition, $N_2$ gas was bubbled for 20 min to avoid oxidation of Fe states in the solution. A pulse electrochemical deposition method was used to prepare NiFe film at both cathodic current density of -10 and -80 mA $cm^{-2}$ (total deposition time 40s; 5s deposition, 10s resting).

### Electrodeposition of NiMo and bifunctional NiFeMo, NiFe, FeMo catalysts

The preparation of NiMo film used in this study was mostly based on the previous reports[4,5]. 0.3 M $NiSO_4{\cdot}6H_2O$, 0.2M $Na_2MoO_4{\cdot}2H_2O$ 0.3M $Na_3C_6H_5O_7{\cdot}2H_2O$ were dissolved in $NH_4OH$. The NiMo film was deposited at a continuous cathodic current density of -160 mA $cm^{-2}$ (in case of NS; -80 mA $cm^{-2}$) for the varied time. The prepared NiMo film was used as a catalyst for electrochemical hydrogen production and as bifunctional catalysts enabling both hydrogen and oxygen production as well. On the basis of electrodeposition conditions of NiMo, we further extended to prepare bifunctional NiFe, FeMo, and NiFeMo catalysts. More detailed deposition conditions are provided in Supplementary Table S2.

## Preparation of photovoltaic driven electrochemical (PV-EC) device

### Lab scale

Small scaled (active area of 0.5 $cm^2$) single base unit of photovoltaic was glued with epoxy (Loctite 9483) onto the bare glass substrate (size of 10 × 10 $cm^2$). Followed by, the glass was securely fixed into a well-sealed polyether ether ketone (PEEK) device holder. In order to decouple PV and EC elements, four wires were connected to contacts of PV and EC by fixing with copper tape. To avoid direct contact between PV and EC, insulating epoxy (Loctite 9483) was covered on the backside of PV before placing EC on there.

### Large (prototype) scale

The in-house designed 3D printing body made from Digital ABS Plus was employed to prepare an upscaled water splitting device. The prepared 4 pieces of anion exchange membrane (fumasep FAA-3-PK-130, with each size of 3 × 8.8 $cm^2$) were mounted with epoxy (Loctite 9492) on the side hydrogen evolved at the 3D printing body. The prepared 8 pieces of bifunctional NiFeMo catalysts (on NS with size of 8.15 cm × 0.8 cm) were clamped into the 3D printing body via a keptone tape and epoxy (Loctite 9483). The contact area of PV was patterned through continuous processes of (1) masking by keptone tape, (2) covering with epoxy (Loctite 9483), and (3) removing keptone tape after 30 min drying. And then, backside of 3D printing device (catalyst side) was covered with 12 × 12 $cm^2$ plexiglass to seal the gap. Subsequently, the patterned PV was integrated into the both catalyst and membrane combined 3D printing body where silver-filled epoxy (Alfa Adhesives; type E10-106) was applied on the exposed contacts of the both PV and EC elements. Finally, in order to protect against some environmental factors and make a stable connection between PV and EC as well, the edge of PV was glued with the epoxy (Loctite 9492) on the front side of 3D printing device. It should be noted that all the works in which epoxy was used were proceeded to the next step after drying for a day in ambient condition.

## Electrochemical measurements

All the measurement were carried out employing potentiostats (Gamry Instruments Reference 600, 1010E, and source measure unit) in 1M potassium hydroxide (KOH) solution in a three (or two) electrode set-up. A platinum mesh and Ag/AgCl were used as counter and a reference electrode, respectively. Reference electrode was calibrated to RHE using the following equation, E (vs. RHE) = E (vs. Ag/AgCl) + 0.21 + 0.059 V × pH. Notably, ohmic voltage loss (*iR* drop) was not compensated in this study since we cannot avoid that loss in our PV-EC device performance (solution resistance of all samples was in error range ~0.2 **Ω**). All the measurements were conducted at a scan rate of 10 mV $s^{-1}$. The integrated PV-EC device was tested using an in-house built sun simulator (Supplementary Fig. 12) at a standard condition (AM 1.5G spectrum; 100 mW $cm^{-2}$, 25 ℃). The evolved gases (hydrogen and oxygen) were collected with an inverted burette bell and fluid level change was monitored over time manually.

## Calculation of solar to hydrogen (STH) efficiency

For the small scale device, the STH efficiency was calculated with below equation 1[6] assuming a faradaic efficiency of 100%.

$$\mathrm{STH} = \left[\frac{I \times (1.23\mathrm{V}) \times \eta_F}{P_{IN} \times A}\right]_{\mathrm{AM\ 1.5G}} \qquad (1)$$

Where $I$ is current of the device, $E_{themo}$ (at 25℃) = 1.23 V, $\eta_F$ is the faradaic efficiency, $P_{IN}$ is incident illumination power density (100 mW $cm^2$), and A is the device area. Here, we have used the photoactive area (0.5 $cm^2$) not considering anode area, because the purpose of the practice in a small device is to gain information how maximum efficiency can be obtained in this structure.

The upscaled standalone and wireless PV-EC device cannot be explored by measuring the current for the calculation of STH efficiency. Instead of monitoring the current, direct gas measurement was conducted to evaluate STH efficiency. The evolution of both hydrogen and oxygen volumes was monitored as a function of operating time as shown in Fig. 4e. The collected gas was converted to STH efficiency (Fig. 4f) using the below equation 2[6].

$$\mathrm{STH} = \left[\frac{(mmol\ H_2/s) \times (237\ \mathrm{kJ/mol})}{P_{IN} \times A}\right]_{\mathrm{AM\ 1.5G}} \qquad (2)$$

Where *mmol $H_2$/s* is the hydrogen production rate, the change in Gibbs free energy per mol of $H_2$ is denoted with 237 kJ/mol (at 25℃), $P_{IN}$ is incident illumination power density (100 mW $cm^2$), and A is the device area. Here, we have used the aperture area (64 $cm^2$) considering the dead (anode and inactive area) area. Notably, all the data are expressed with a change in volume of the gas assuming 100% purity of both $H_2$ and $O_2$.

## Spectroscopy characterization

The surface morphologies of the prepared catalysts were investigated by using a scanning electron microscope (SEM, FEI Magellan 400) at accelerating voltages of 20 kV. An energy dispersive X-ray spectrometer (EDX, Oxford X-Max 80mm²) was employed for elemental mapping analysis on the samples. Grazing incidence X-ray diffraction (GIXRD) was conducted using Empyrean (Malvern Panalytical GmbH) for the investigation of the surface state of the prepared samples. The data were collected using CuKα radiation (45 kV/ 40 mA). X-ray photoelectron spectroscopy (XPS) measurements were performed with XM1000 Al Ka x-ray source (300 W) at the JOSEPH cluster system in the research center Juelich. To examine the three-dimensional depth-profiling view of the composition on the prepared samples, the time-of-flight secondary ion mass spectrometry (TOF-SIMS) analysis was conducted (ION-TOF, Germany). A 25-keV Bi1+ beam was employed to analyze the elements, and the depth profiling was analyzed with a 10-keV beam of $Cs^+$. Raman spectroscopy (Alpha 300S, WITec) with laser-excitation energy of 532 nm (Nd;Yag laser) was used to understand the structure of the prepared samples.

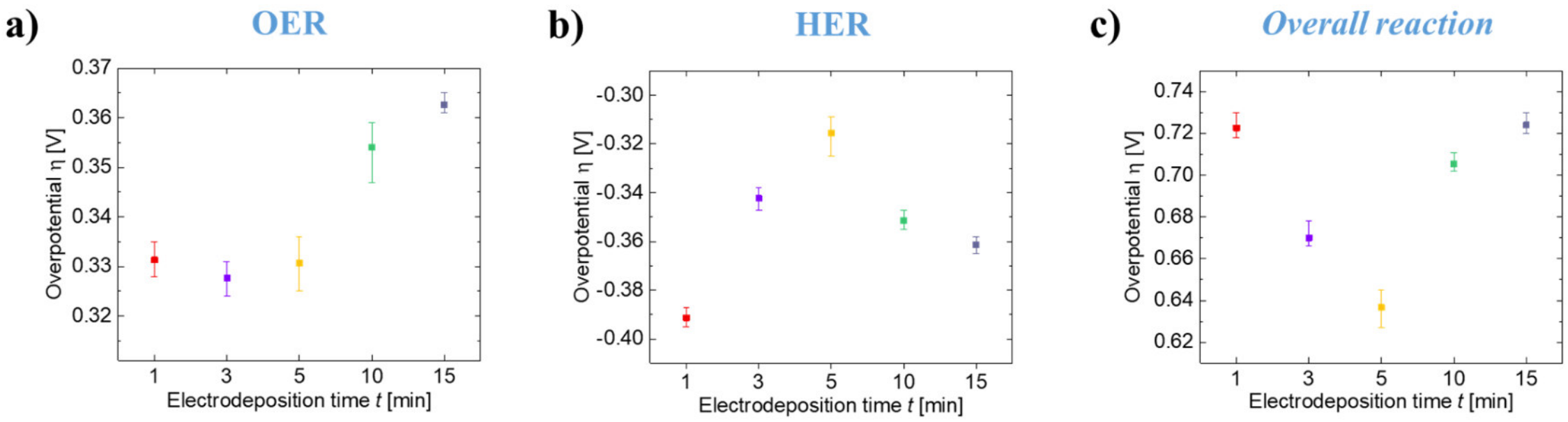

**Supplementary Figure 1.** Water splitting performance of NiFeMo bifunctional catalyst as a function of electrodeposition time; (a) OER (b) HER, and (c) overall reaction. The overpotential of catalysts was measured at a current density of 50 mA $cm^{-2}$. Three samples (with size of 1 x 3 $cm^2$) were measured for each condition.

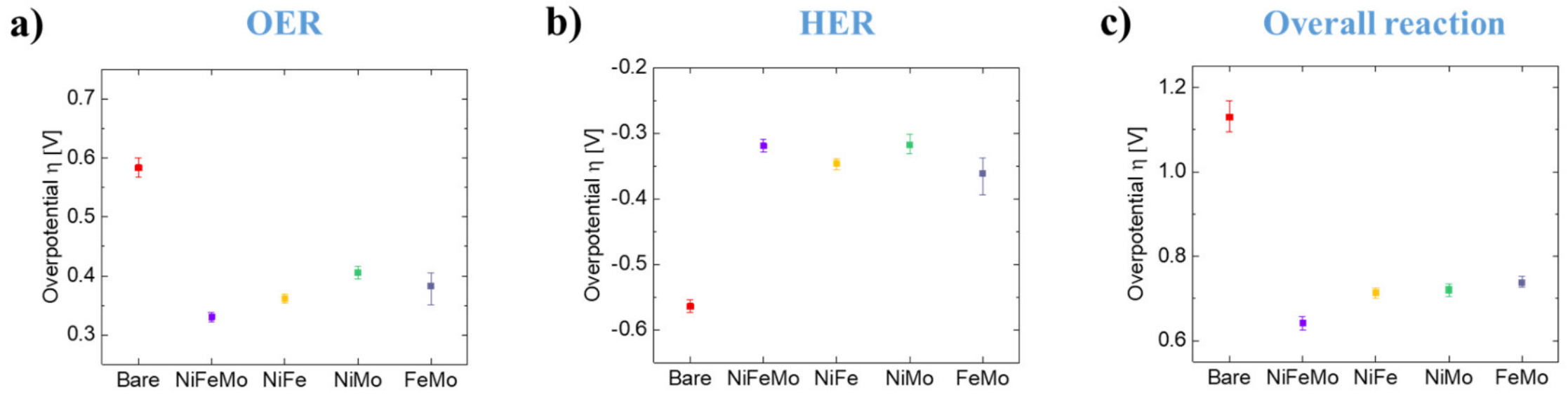

**Supplementary Figure 2.** Water splitting performance with different catalysts; (a) OER (b) HER, and (c) overall reaction. The overpotential of catalysts was measured at a current density of 50 mA $cm^{-2}$. Three samples (with size of 1 x 3 $cm^2$) were measured for each condition.

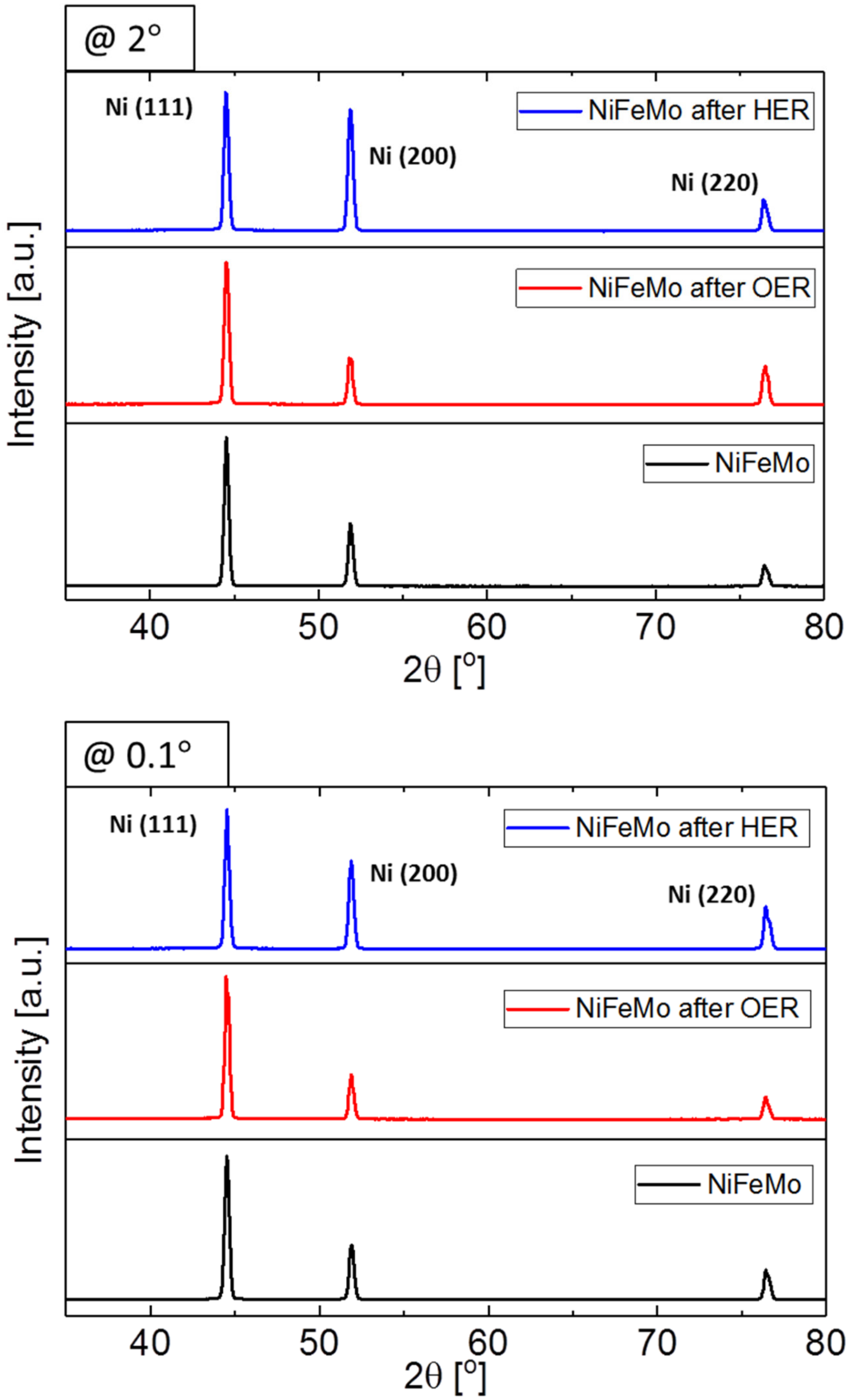

**Supplementary Figure 3.** Grazing incidence X-ray diffraction (GIXRD) patterns of as-prepared NiFeMo and the samples after both reactions (OER and HER) with different incident angles (2° and 0.1°).

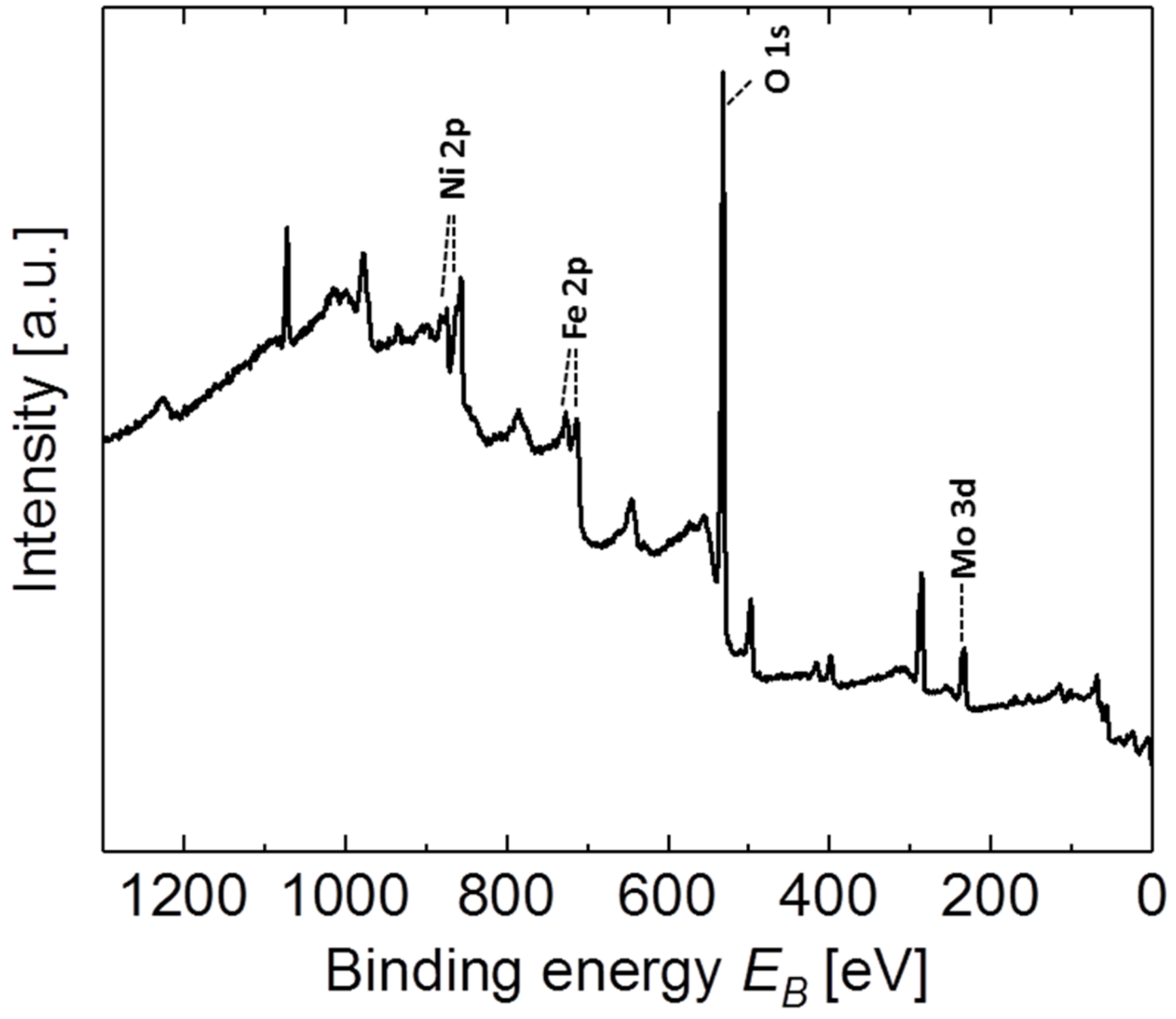

**Supplementary Figure 4.** XPS elemental survey on as-prepared NiFeMo.

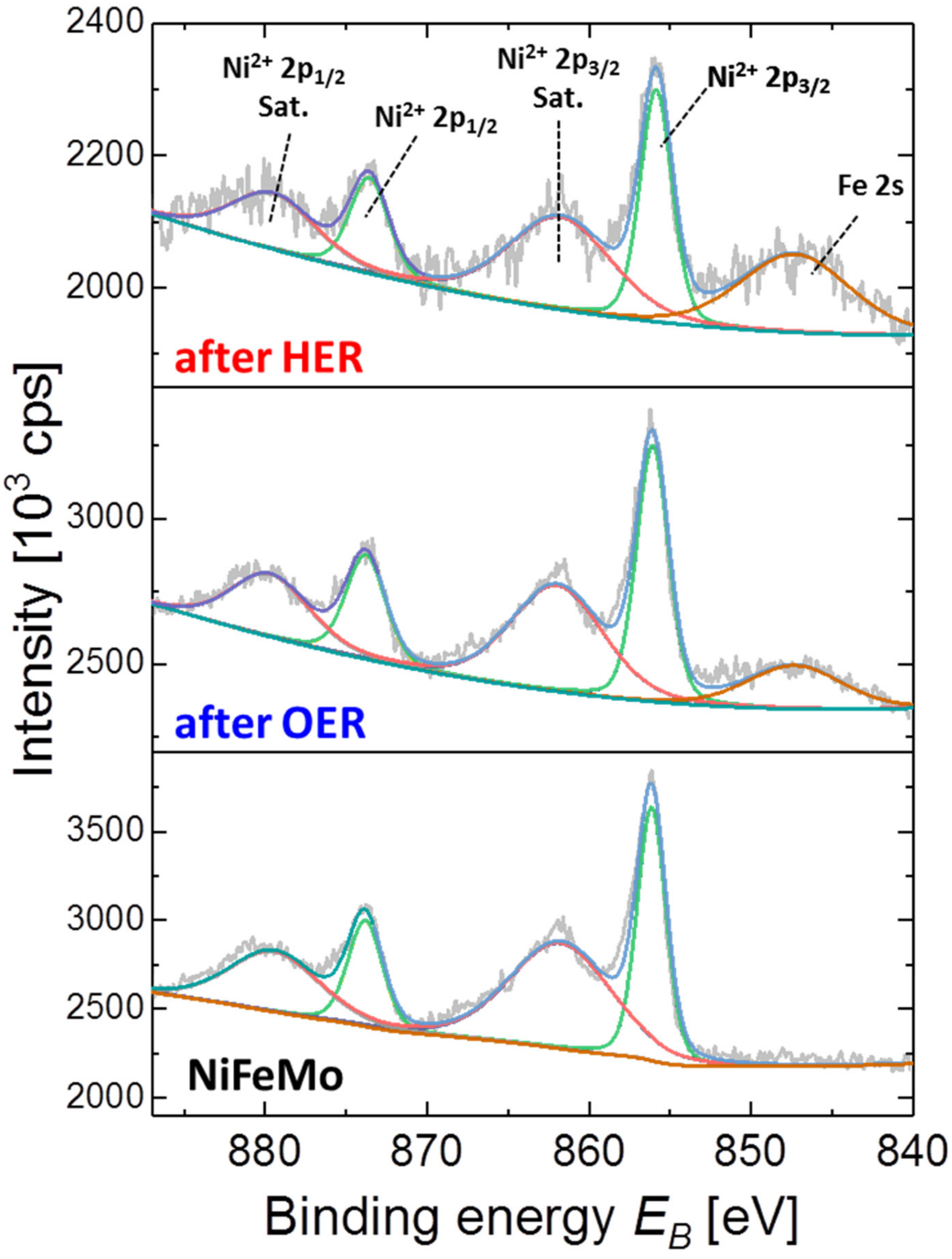

**Supplementary Figure 5.** Ni 2p XPS peaks of for NiFeMo bifunctional catalysts before and after reaction.

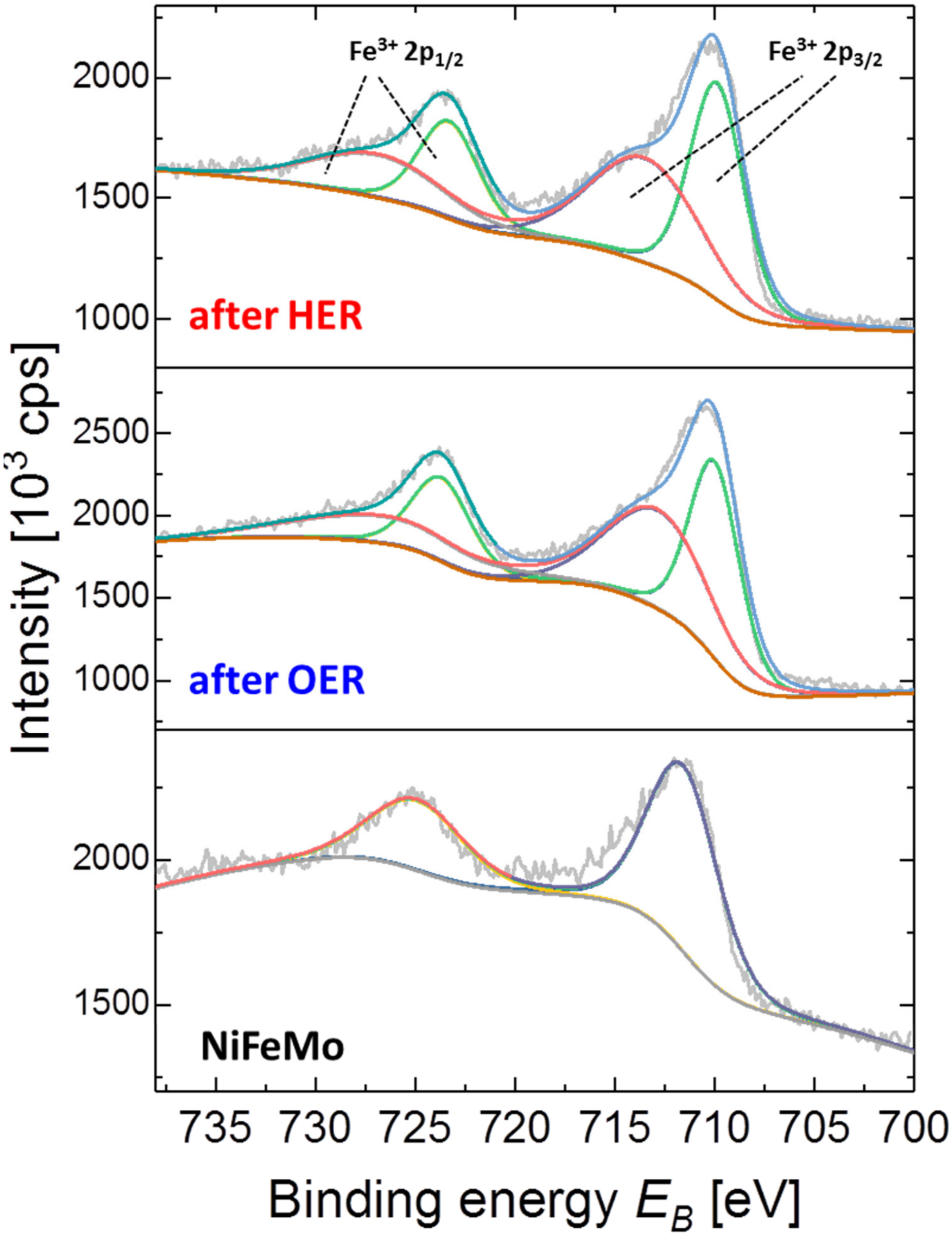

**Supplementary Figure 6.** Fe 2p XPS peaks of for NiFeMo bifunctional catalysts before and after reaction.

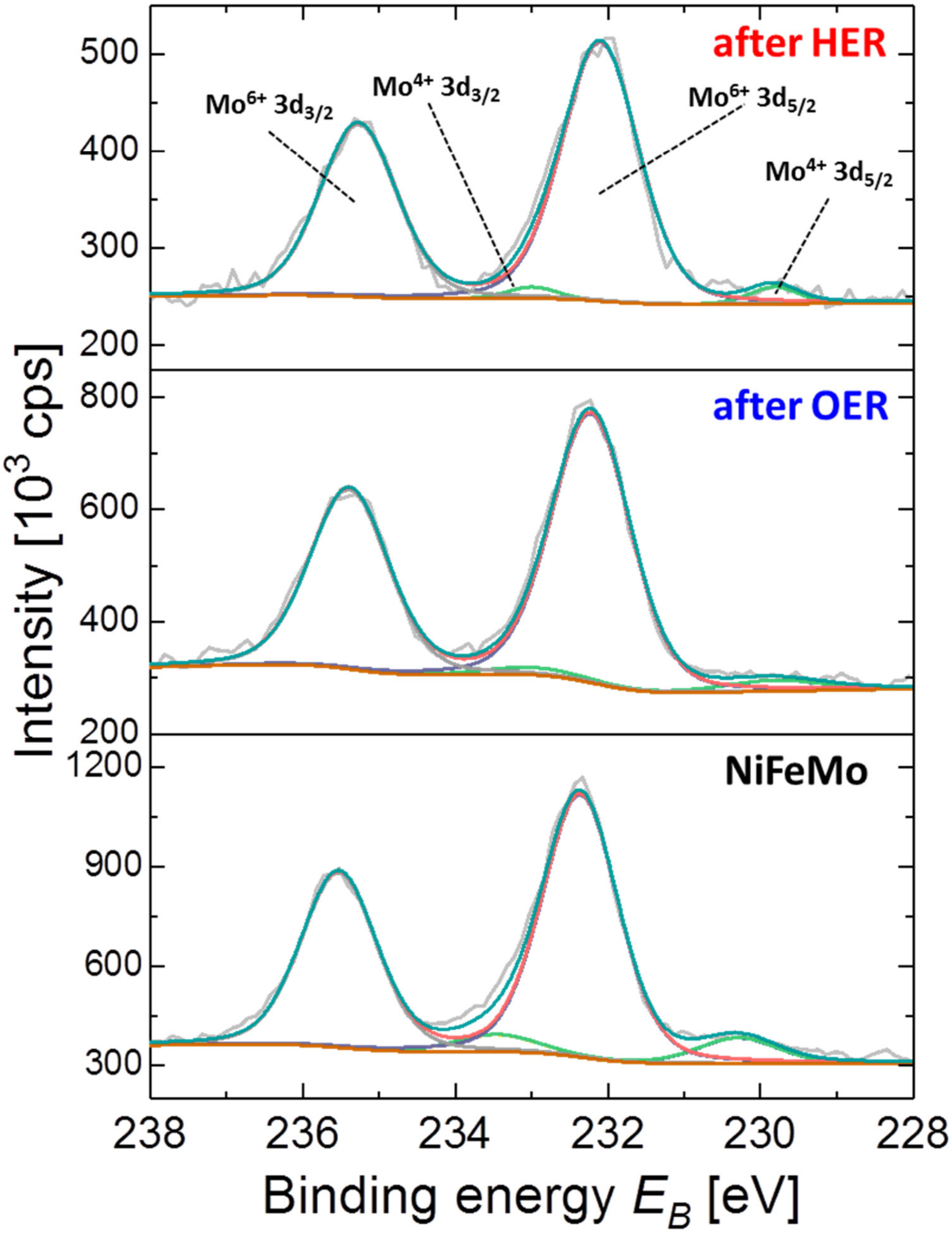

**Supplementary Figure 7.** Mo 3d XPS peaks of for NiFeMo bifunctional catalysts before and after reaction.

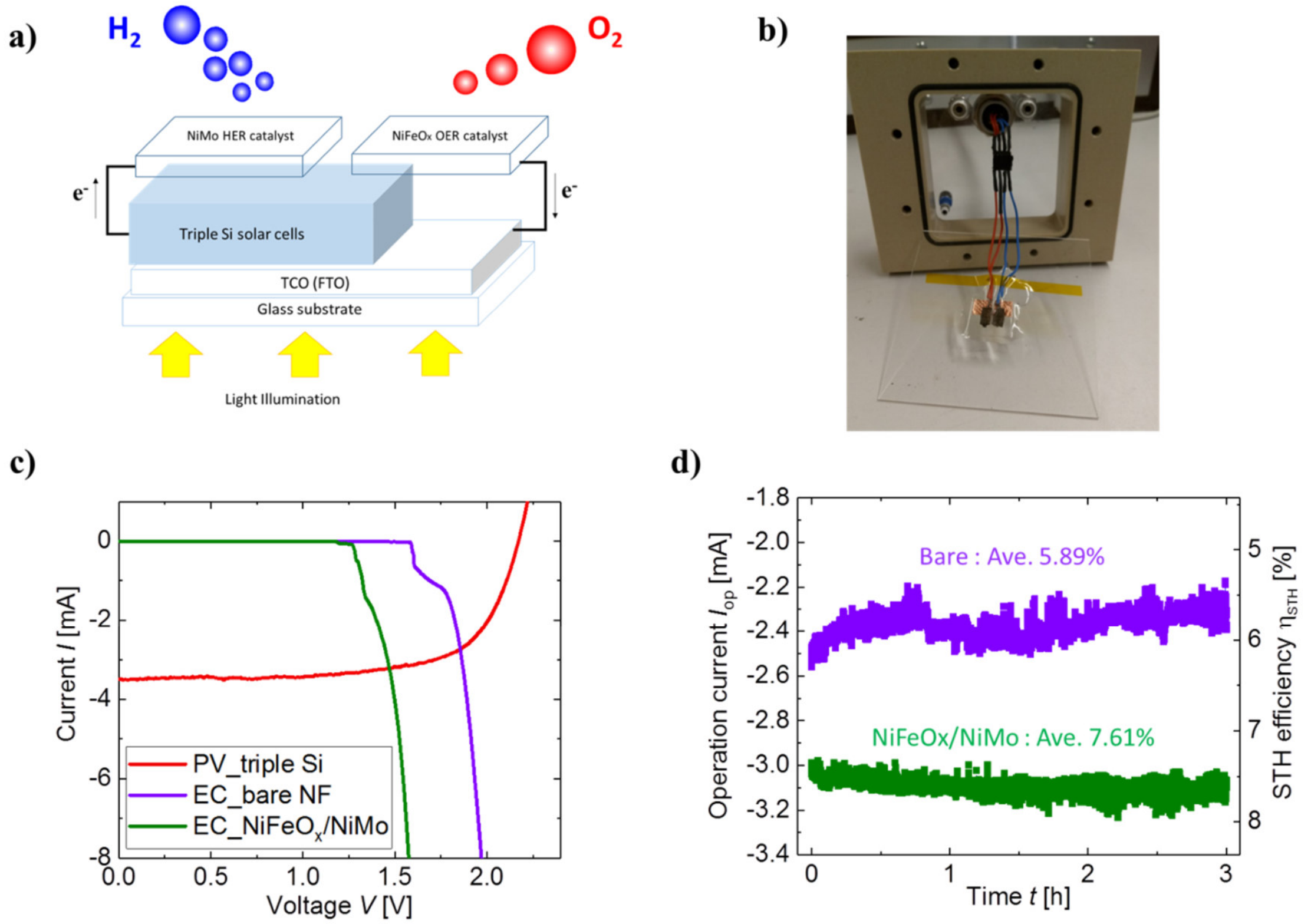

**Supplementary Figure 8.** (a) Schematic drawing and (b) Photograph of small area PV-EC device. As parts of the PV-EC device, triple junction a-Si:H/a-Si:H/μc-Si:H solar cell and benchmark water splitting catalysts ($NiFeO_x$ for OER and NiMo for HER) were used as PV and EC, respectively. (c) The *I*-*V* curves of individual PV and EC. The intersection where PV and EC curves meet can be regarded as the working point for the spontaneous water splitting. (d) Operation current of combined PV-EC device as a function of time under bias-free condition.

a)

b)

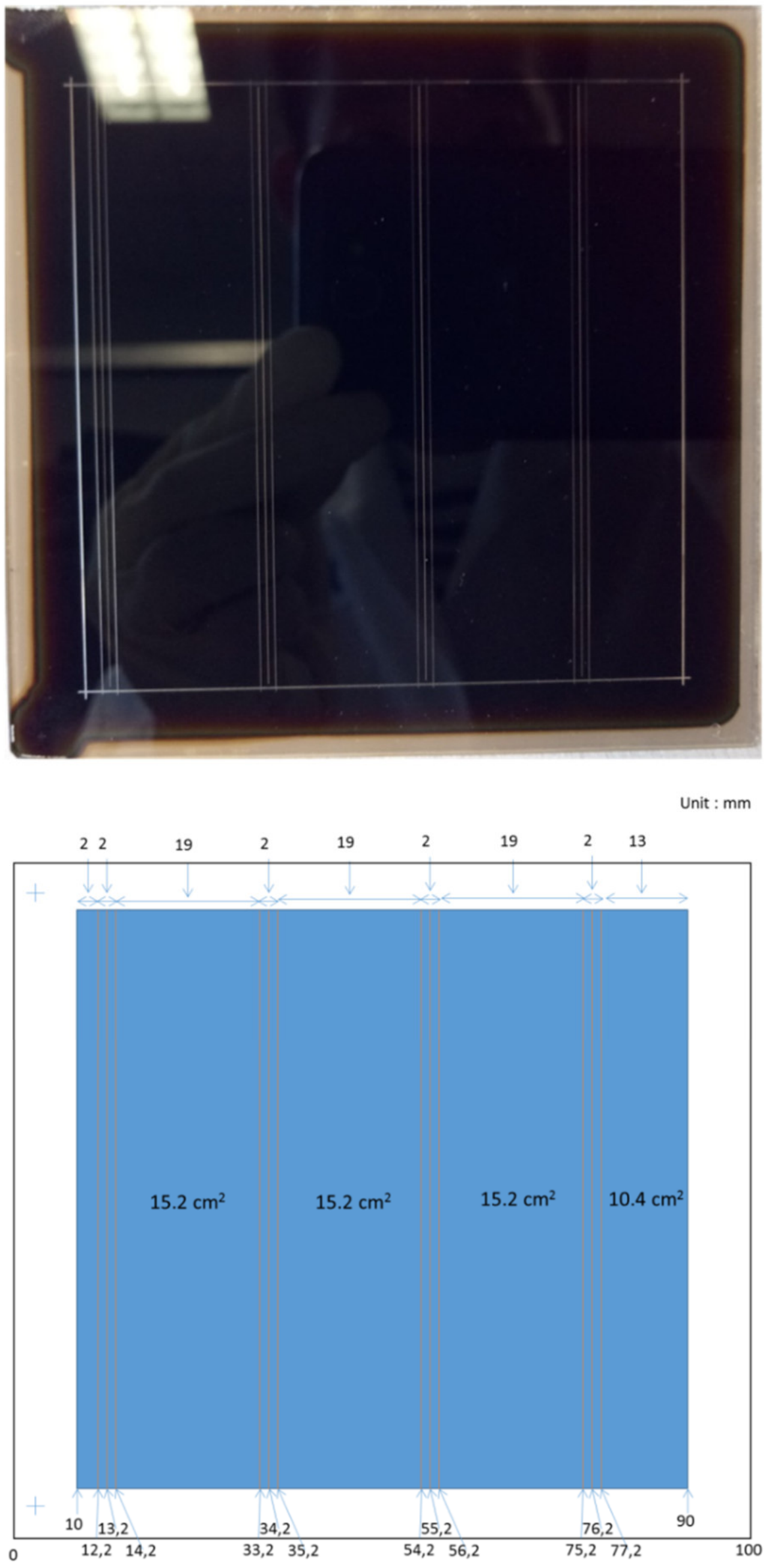

**Supplementary Figure 9.** (a) photograph and (b) sketch of the detailed dimensions of the large scale triple junction Si solar cell used in this study.

The triple junction Si solar cell was prepared on 10 x 10 $cm^2$ substrate. The laser ablation was employed to expose the front contact for the connection of OER catalyst. The total aperture area was 64 $cm^2$ with an active area of 56 $cm^2$.

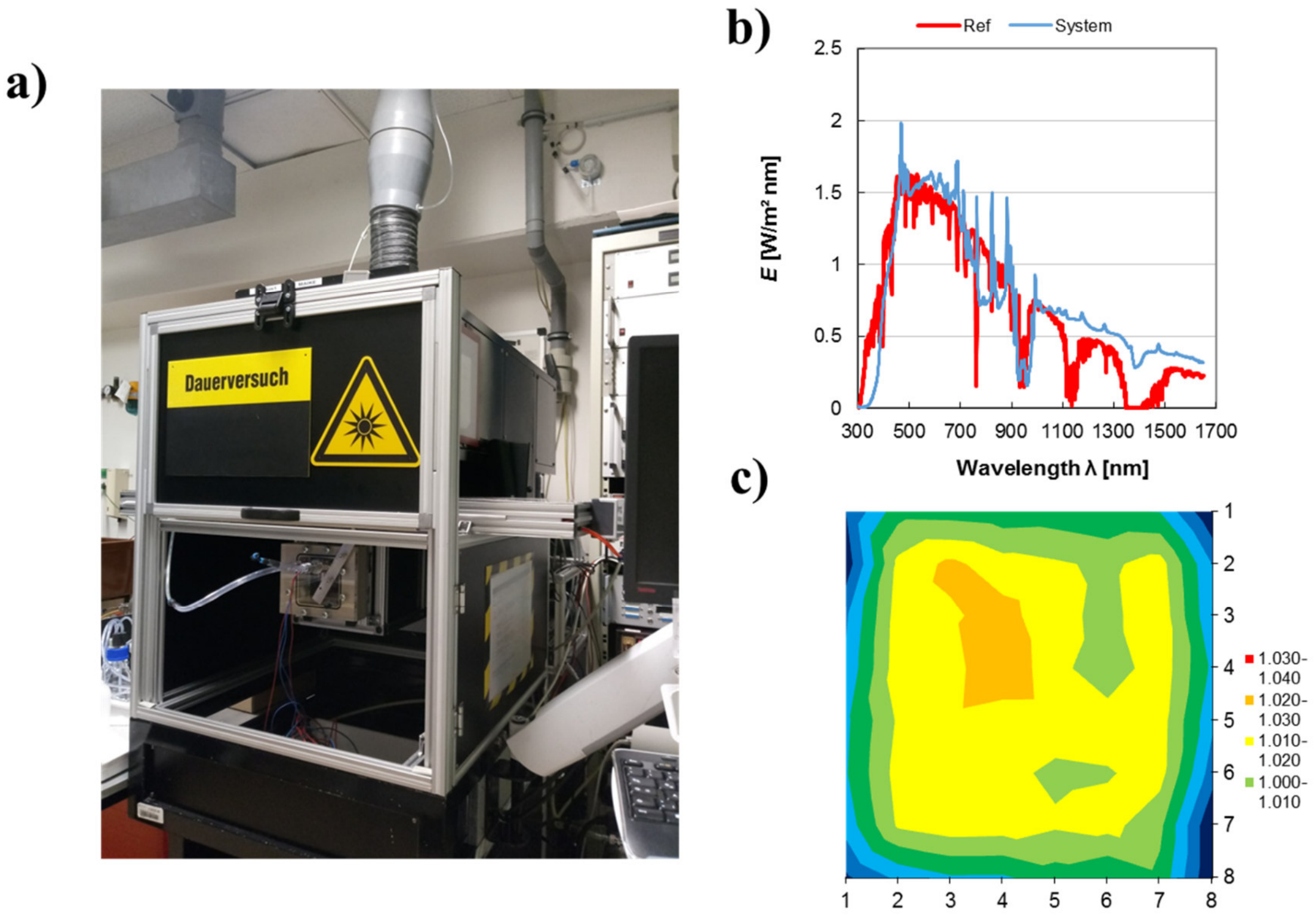

**Supplementary Figure 10.** (a) Photograph of the measurement in-house built setup. The measurement was conducted in the black box where large area sun simulator was equipped. (b) The plot of spectral power density as a function of the wavelength of sun simulator in the setup system. It can be seen that both plots are not significantly different, especially the photon absorption region of triple junction Si solar cells (400 nm to 1100 nm). (c) Mapping image of light illumination at the operating condition of sun simulator. The value of homogeneity (± 1.85%) was obtained with an 8 x 8 array diodes system. The total distance between the edges of diodes is 86.4 mm.

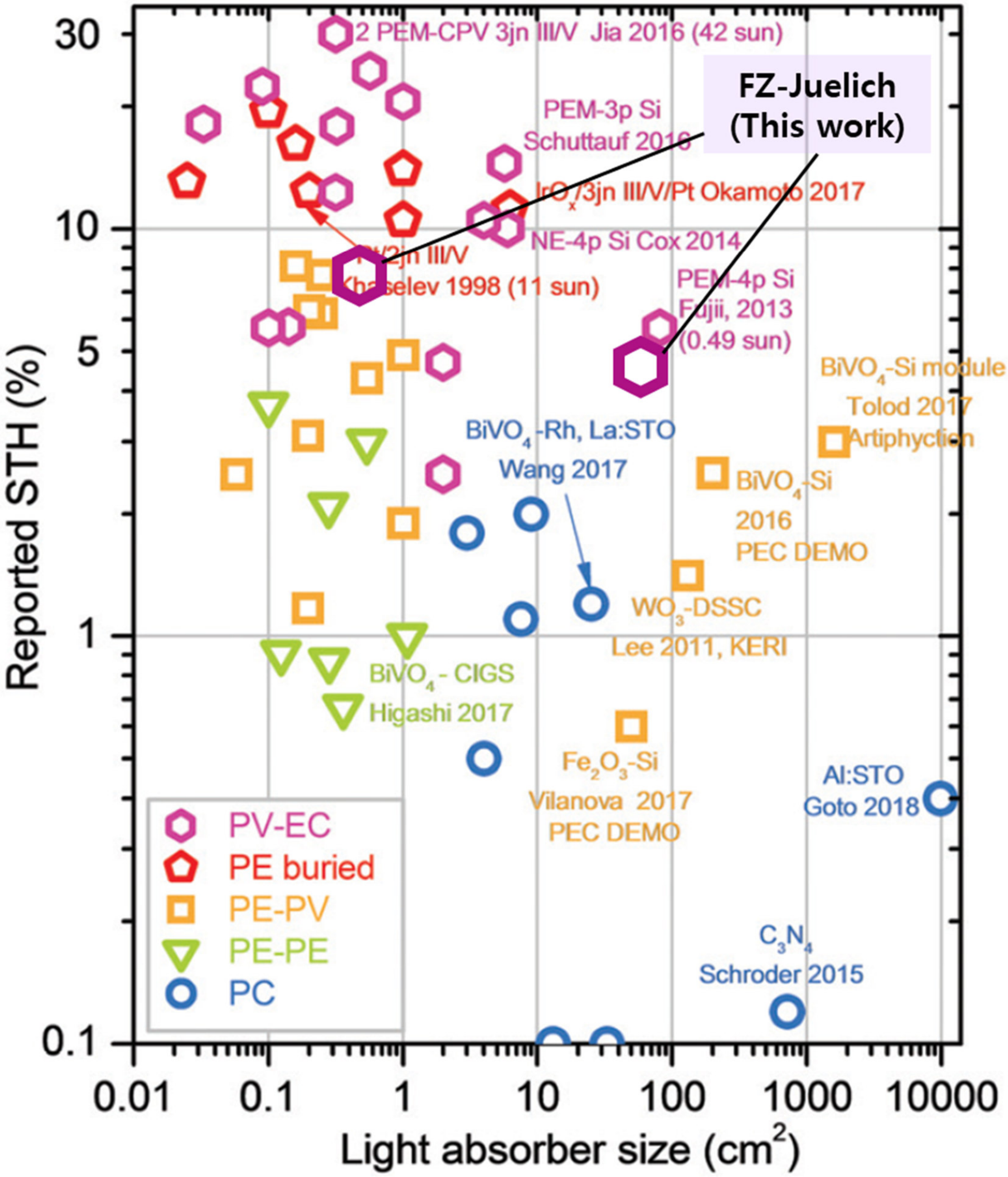

**Supplementary Figure 11.** A compilation of reported STH efficiencies as a function of light absorber size. The graph was compiled based on the data from Ref. 7 (Fig. 21c).

PV-EC: photovoltaic-electrolysis, PE buried: photoelectrode buried, PE-PV: photoelectrode-photovoltaic, at least one side of the device directly contacts water forming a semiconductor liquid junction, PE-PE: photoelectrode-photoelectrode, PC: photocatalyst

**Supplementary Table 1.** Average PV parameter values (samples in Figure 1b) in batches of 4 cells are given in the table. 'Sample A' indicates triple junction a-Si:H/a-Si:H/μc-Si:H solar cell without intermediate reflecting layer. 'Sample B' indicates triple junction a-Si:H/a-Si:H/μc-Si:H solar cell with n-type μc-$SiO_x$:H intermediate reflecting layer.

| | **Efficiency (%)** | $J_{sc}$ **(mA cm$^{-2}$)** | $V_{oc}$ ***(V)*** | **FF** |
|---|---|---|---|---|
| **sample A** | 9.47±0.099 | 6.41±0.044 | 2.14±0.013 | 69.20±0.641 |
| **sample B** | 10.83±0.162 | 7.29±0.010 | 2.16±0.014 | 68.76±0.637 |

**Supplementary Table 2**. The solution composition of the bifunctional catalysts. All the precursors were dissolved in $NH_4OH$. The deposited film was prepared at a continuous cathodic current density of -160 mA $cm^{-2}$ (for -80 mA $cm^{-2}$ for NS) for the varied time.

| | | | | |
|---|---|---|---|---|
| **NiMo** | 0.3M $NiSO_4 \cdot 6H_2O$ | | 0.2M $Na_2MoO_4 \cdot 2H_2O$ | 0.3M $Na_3C_6H_5O_7 \cdot 2H_2O$ |
| **NiFe** | 2.4M $NiSO_4 \cdot 6H_2O$ | 0.6M $FeSO_4 \cdot 6H_2O$ | | 0.3M $Na_3C_6H_5O_7 \cdot 2H_2O$ |
| **FeMo** | | 0.3M $FeSO_4 \cdot 6H_2O$ | 0.2M $Na_2MoO_4 \cdot 2H_2O$ | 0.3M $Na_3C_6H_5O_7 \cdot 2H_2O$ |
| **NiFeMo** | 2.4M $NiSO_4 \cdot 6H_2O$ | 0.6M $FeSO_4 \cdot 6H_2O$ | 0.2M $Na_2MoO_4 \cdot 2H_2O$ | 0.3M $Na_3C_6H_5O_7 \cdot 2H_2O$ |

**Supplementary Table 3.** Comparison table of state-of-the-art results in terms of bias-free photovoltaic driven water splitting system previously reported.

| Device configuration | PV (or PA) | PCE (PV) | Catalysts (OER/HER)/ Electrolyte | STH $\eta$/time | Product Separation /Anal. method | Scale ($cm^2$) | Ref. |
|---|---|---|---|---|---|---|---|
| 2-junction tandem cell | GaInP/GaAs | 13.86% | Ni/NiMo /1M KOH | 8.6% (with gas) at 1sun /2h | O /both quan. And qual. | 1$cm^2$ w/ac | 8 |
| 2-junction tandem cell | GaInP/GaInAs | N/A | $RuO_2$/Rh 1M $HClO_4$ | 19.3% (with $J_{op}$) at 1sun /20h (with 0.5M $KH_2PO_4$ /$K_2HPO_4$, 17% loss) | O /quan. | 0.1-0.3 $cm^2$ | 9 |
| 3-junction PV module | GaInP/GaAs/ GaInNAs(Sb) | 38.9% | Ir/Pt /$H_2SO_4$-impregnated Nafion | 31.2% (with $J_{op}$) at 42sun, 80℃ /48h (~10% loss) | O /quan. | 0.316 $cm^2$ w/ac | 10 |
| 3-junction PV cell | TF-Si (a-Si:H/a-Si:H /μc-Si:H) | 13.6% | $RuO_2$/Pt /1M KOH | 9.5% (with $J_{op}$) at 1sun, R.T. / >4h (with 0.1M KOH) | X /quan. | 0.5$cm^2$ w/ac | 1 |
| 3-junction PV cell | TF-Si (3jn-a-Si) | 7.7% | Co-OEC /NiMoZn /1M KBi | 2.5% (with $J_{op}$) at 1sun / >24h (~20% loss) | X /quan. | N/A | 11 |
| 4-lateral series connected PV module | c-Si | N/A | Pt–Pt /water | 6.18% (with $J_{op}$) /~18m (~5% loss) | O /quan. | 81.2$cm^2$ w/ap | 12 |
| 4-lateral series connected PV module | c-Si | 16% | Ni-Bi/NiMoZn /0.5 M KBi | 10% (with $J_{op}$) /~168m (~3% loss) | X /qual. | 6$cm^2$ | 13 |
| 3-lateral series connected PV module | c-Si | 20.6% | NiF/NF /1M KOH | 14.2% (with $J_{op}$) /100h | X /(N/A) | 5.7cm$^2$ w/ac | 14 |
| 3-lateral series connected PV module | c-Si | 20.57% | NiFe/NiMo /KOH-impregnated PVA | 15.1% (with $J_{op}$) /1500s | O /qual. (w/o PV) | 3.54cm$^2$ w/ac | 15 |
| 3-lateral series connected PV module | CIGS | 17% | Pt/Pt /3M $H_2SO_4$ | 10.5~11% (with $J_{op}$) at 1sun /27 | X /both quan. and qual. | N/A less than 10cm$^2$ | 16 |
| 2-lateral series connected PV module | Porphyrin dye (DSSC) | 12.3% (0.141 cm$^2$) | Pt/Pt /1M NaOH | 5.75% (with $I$-$V$) at 1sun /5h (at 0.5sun, ~10% loss) | X /qual. | 0.282 cm$^2$ w/ac | 17 |
| 2-lateral series connected 2-junction tandem module | DSSC/CIGS | 7.33% | SUS304/Pt /1M KOH | 4.65% (with $J_{op}$) at 1sun /5m | O /qual. | 0.5cm$^2$ w/ac | 18 |
| 3-junction PV cell | PCDTBT:PCBM[70]/PMDPP3T PCBM[60]/PMDPP3T PCBM[60] (OSC) | 6.7% (0.0676 cm$^2$) | $RuO_2$–$RuO_2$, /1M KOH | 3.6% /20m | X /(N/A) | 0.0676 cm$^2$ w/ac | 19 |
| 2-lateral series connected PV module | Lead halide perovskite | 15.7% | $NiFe(OH)_2$/ $NiFe(OH)_2$ (on NF) /1M NaOH | 12.3% (with $J_{op}$) /4h | X /qual. | 0.318 cm$^2$ w/ac | 20 |
| 2-lateral series connected PV module | Lead halide perovskite | N/A | NiFe LDH–CoP /1M KOH (BPM) 0.5M $H_2SO_4$ | 12.7% (with $J_{op}$) at 1sun /16h (~29% loss) | X /qual. | 0.32cm$^2$ | 21 |
| 2-junction tandem cell | Lead halide perovskite /c-Si | 23.1% (0.1875 $cm^2$) | NiFe(O)OH /$Ni_4Mo$ /1M NaOH | 17.52% (with $J_{op}$) at 1sun /~6.7m | X /(N/A) | 0.1875 $cm^2$ | 22 |
| 13 neighbouring base unit consisting of 2-lateral series connected 2-junction tandem PV | TF-Si (a-Si:H/ μc-Si:H) | 11.49% (1$cm^2$) | NF/NF /1M KOH | ~3.9% (with gas) at 1sun, R.T. /3h | X /quan. | 64$cm^2$ w/ap | 3 |
| 3-junction PV module | TF-Si (a-Si:H/a-Si:H /μc-Si:H) | 10.2% (1$cm^2$) | $IrO_x$/Pt /1M KOH | ~4.8% (with $J_{op}$) at 1sun, R.T. /80m | O /quan. | 64$cm^2$ w/ap | 2 |
| 3-junction PV module | TF-Si (a-Si:H/a-Si:H /μc-Si:H) | 7.7% (64$cm^2$) | $NiFeO_x$/NiMo /1M KOH | ~5.1% (with *I-V*) at 1sun, R.T. /10m | X /(N/A) | 64$cm^2$ w/ap | 23 |
| 3-junction PV cell | TF-Si (a-Si:H/a-Si:H /μc-Si:H) | 10.78% (2.25$cm^2$) | NiFeMo/ NiFeMo (on NF) /1M KOH | ~7.72% (with $J_{op}$) at 1sun, R.T. /100h (17% loss) | X /(N/A) | 0.5$cm^2$ w/ac | **This work** |
| 7 neighbouring base unit consisting of 3-junction PV module | TF-Si (a-Si:H/a-Si:H /μc-Si:H) | N/A | NiFeMo/ NiFeMo /1M KOH | ~4.67% (with gas) at 1sun, R.T. /30m | O /quan. | 64$cm^2$ w/ap | **This work** |

PCE: power conversion efficiency, PV: photovoltaic, PA: photoabsorber, STH $\eta$: solar to hydrogen efficiency, Anal.: analysis, Ref.: reference, quan.:gas products were quantitatively measured, qual.: gas products were qualitatively measured, w/ap: with aperture area, w/ac: with active area, N/A: not available, TF-Si: thin film silicon, c-Si: crystallized silicon, CIGS: $Cu(In_xGa_{1-x})(S_ySe_{1-y})_2$, DSSC: dye-sensitized solar cell, OSC: organic solar cell, $J_{op}$: operation current density, *I-V*: current voltage curve, OEC: oxygen-evolving complex, NF: nickel foam, LDH: layered double hydroxides, BPM: bipolar membrane, (a-Si:H/μc-Si:H): hydrogenated amouphous and micro-crystalline Si, R.T.: room temperature, PVA: poly(vinyl alcohol).

## Reference

1 Urbain, F. *et al.* Multijunction Si photocathodes with tunable photovoltages from 2.0 V to 2.8 V for light induced water splitting. *Energy Environ. Sci.* **9**, 145-154, (2016).

2 Becker, J. P. *et al.* A modular device for large area integrated photoelectrochemical water-splitting as a versatile tool to evaluate photoabsorbers and catalysts. *J. Mater. Chem. A* **5**, 4818-4826, (2017).

3 Turan, B. *et al.* Upscaling of integrated photoelectrochemical water-splitting devices to large areas. *Nat. Commun.* **7**, 12681, (2016).

4 McCrory, C. C. L. *et al.* Benchmarking Hydrogen Evolving Reaction and Oxygen Evolving Reaction Electrocatalysts for Solar Water Splitting Devices. *J. Am. Chem. Soc.* **137**, 4347-4357, (2015).

5 Fan, C., Piron, D. L., Sleb, A. & Paradis, P. Study of Electrodeposited Nickel-Molybdenum, Nickel-Tungsten, Cobalt-Molybdenum, and Cobalt-Tungsten as Hydrogen Electrodes in Alkaline Water Electrolysis. *J. Electrochem. Soc.* **141**, 382-387, (1994).

6 Chen, Z. *et al.* Accelerating materials development for photoelectrochemical hydrogen production: Standards for methods, definitions, and reporting protocols. *J. Mater. Res.* **25**, 3-16, (2011).

7 Kim, J. H., Hansora, D., Sharma, P., Jang, J.-W. & Lee, J. S. Toward practical solar hydrogen production – an artificial photosynthetic leaf-to-farm challenge. *Chem. Soc. Rev.* **48**, 1908-1971, (2019).

8 Verlage, E. *et al.* A monolithically integrated, intrinsically safe, 10% efficient, solar-driven water-splitting system based on active, stable earth-abundant electrocatalysts in conjunction with tandem III–V light absorbers protected by amorphous TiO2 films. *Energy Environ. Sci.* **8**, 3166-3172, (2015).

9 Cheng, W.-H. *et al.* Monolithic Photoelectrochemical Device for Direct Water Splitting with 19% Efficiency. *ACS Energy Lett.* **3**, 1795-1800, (2018).

10 Jia, J. *et al.* Solar water splitting by photovoltaic-electrolysis with a solar-to-hydrogen efficiency over 30%. *Nat. Commun.* **7**, 13237, (2016).

11 Reece, S. Y. *et al.* Wireless Solar Water Splitting Using Silicon-Based Semiconductors and Earth-Abundant Catalysts. *Science* **334**, 645-648, (2011).

12 Fujii, K. *et al.* Characteristics of hydrogen generation from water splitting by polymer electrolyte electrochemical cell directly connected with concentrated photovoltaic cell. *Int. J. Hydrog. Energy* **38**, 14424-14432, (2013).

13 Cox, C. R., Lee, J. Z., Nocera, D. G. & Buonassisi, T. Ten-percent solar-to-fuel conversion with nonprecious materials. *Proc. Natl. Acad. Sci. U.S.A.* **111**, 14057, (2014).

14 Schüttauf, J.-W. *et al.* Solar-to-Hydrogen Production at 14.2% Efficiency with Silicon Photovoltaics and Earth-Abundant Electrocatalysts. *J. Electrochem. Soc.* **163**, F1177-F1181, (2016).

15 Heremans, G. *et al.* Vapor-fed solar hydrogen production exceeding 15% efficiency using earth abundant catalysts and anion exchange membrane. *Sustain. Energy Fuels* **1**, 2061-2065, (2017).

16 Jacobsson, T. J., Fjällström, V., Sahlberg, M., Edoff, M. & Edvinsson, T. A monolithic device for solar water splitting based on series interconnected thin film absorbers reaching over 10% solar-to-hydrogen efficiency. *Energy Environ. Sci.* **6**, 3676-3683, (2013).

17 Kang, S. H. *et al.* Porphyrin Sensitizers with Donor Structural Engineering for Superior Performance Dye-Sensitized Solar Cells and Tandem Solar Cells for Water Splitting Applications. *Adv. Energy Mater.* **7**, 1602117, (2017).

18 Chae, S. Y., Park, S. J., Joo, O.-S., Min, B. K. & Hwang, Y. J. Spontaneous solar water splitting by DSSC/CIGS tandem solar cells. *Sol. Energy* **135**, 821-826, (2016).

19 Esiner, S. *et al.* Photoelectrochemical water splitting in an organic artificial leaf. *J. Mater. Chem. A* **3**, 23936-23945, (2015).

20 Luo, J. *et al.* Water photolysis at 12.3% efficiency via perovskite photovoltaics and Earth-abundant catalysts. *Science* **345**, 1593-1596, (2014).

21 Luo, J. *et al.* Bipolar Membrane-Assisted Solar Water Splitting in Optimal pH. *Adv.*

*Energy Mater.* **6**, 1600100, (2016).
22 Park, H. *et al.* Water Splitting Exceeding 17% Solar-to-Hydrogen Conversion Efficiency Using Solution-Processed Ni-Based Electrocatalysts and Perovskite/Si Tandem Solar Cell. *ACS Appl. Mater. Interfaces* **11**, 33835-33843, (2019).
23 Welter, K. *et al.* Catalysts from earth abundant materials in a scalable, stand-alone photovoltaic-electrochemical module for solar water splitting. *J. Mater. Chem. A* **6**, 15968-15976, (2018).